\documentstyle[preprint,aps,epsfig]{revtex}
\evensidemargin=\oddsidemargin  

\def\today{\number\day
	   \space\ifcase\month\or
	     January\or February\or March\or April\or May\or June\or
	     July\or August\or September\or October\or November\or 
             December\fi
	   \space\number\year}
\def\thisday{July 15, 1998}
\def\be{\begin{equation}}
\def\ee{\end{equation}}
\def\bea{\begin{eqnarray}}
\def\eea{\end{eqnarray}}

\def\ifb{ {\rm fb}^{-1} }

\def\to{\rightarrow}

\def\dis{\displaystyle}
\def\f{\frac}
\def\Psibar{\bar{\Psi}}
\def\ppbar{p\bar{p}}
\def\bbar{\bar{b}}
\def\tbar{\bar{t}}
\def\phibb{\phi b \bar{b}}

\def\bbbb{b \bar{b} b \bar{b}}
\def\bbjj{b \bar{b} j j}
\def\Zbb{Z b \bar{b}}
\def\ifb{${\rm fb}^{-1}$}
\newcommand{\tanb}{\tan\hspace*{-0.9mm}\beta}
\newcommand{\cotb}{\cot\beta}
\newcommand{\sinb}{\sin\beta}
\newcommand{\cosb}{\cos\beta}
\newcommand{\cosa}{\cos\alpha}
\newcommand{\sina}{\sin\alpha}
\newcommand{\cosba}{\cos (\beta -\alpha)}
\newcommand{\sinba}{\sin (\beta -\alpha)}

\newcommand{\lae}{\stackrel{<}{\sim}}
\newcommand{\gae}{\stackrel{>}{\sim}}

\begin{document}

\thispagestyle{empty}
\hspace*{-0.65cm} hep-ph/9807349      \hfill  MSUHEP-80515 \\
\thisday      \hfill  FERMILAB-PUB-98/182-T  \\
Oct.~30, 1998 (Final Version)  \hfill        ANL-HEP-PR-98-55

\vspace{0.5cm} 

\begin{center}
{ {\large {\bf
Probing Higgs Bosons with Large Bottom Yukawa Coupling \\[0.18cm]
at Hadron Colliders}}} \\[1.2cm]

\centerline{ 
{\sc C.~Bal\'azs}$^{1,2}$,~~{\sc J.L.~Diaz-Cruz}$^3$,~~ 
{\sc H.--J. He}$^{1,2}$,~~{\sc T.~Tait}$^{1,4}$,~~ {\sc C.--P.~Yuan}$^1$
}

$^1$~{\it Michigan State University, 
East Lansing, Michigan 48824, USA} \\
$^2$~{\it Fermi National Accelerator Laboratory, Batavia, 
Illinois 60510, USA} \\
$^3$~{\it Instituto de Fisica, BUAP, 72570 Puebla, Pue, Mexico} \\
$^4$~{\it Argonne National Laboratory, Illinois 60439, USA}

\begin{abstract}
\noindent 
The small mass of the bottom quark, relative to 
its weak isospin partner, the top quark,
makes the bottom an effective probe of new 
physics in Higgs and top sectors.
We study the Higgs boson production associated with
bottom quarks, $p\bar{p}/pp \to \phibb \to \bbbb$,
at the Fermilab Tevatron and the CERN LHC. 
We find that strong and model-independent constraints on the
size of the $\phi$-$b$-$\bar b$ coupling can be obtained for a wide
range of Higgs boson masses.  Their implications for the
composite Higgs models with strong dynamics associated with the
third family quarks (such as the top-condensate/topcolor models
with naturally large bottom Yukawa couplings),
and for the supersymmetric models
with large $\tan\beta$, are analyzed. 
We conclude that the Tevatron 
and the LHC can put stringent bounds on these models,
if the $\phibb$ signal is not found. 
\\[0.4cm]
PACS number(s): 14.80.Cp, 12.38.Bx, 12.60.Jv, 14.65.Fy
~~~~~~~~~~~~~~~~\\
[3mm]
{\it Physical Review D (1998),} in press.~~~~~~~~~~~~~~~~~~~
\end{abstract}
\end{center}

\newpage
\setcounter{footnote}{0}
\renewcommand{\thefootnote}{\arabic{footnote}}

\section{Introduction}

A major task for all future high energy colliders is to
determine the electroweak symmetry breaking (EWSB) mechanism 
for generating the $(W^\pm ,Z^0)$ masses and the mechanism for the 
fermion mass generation \cite{hhunt}. 
Whether the two mechanisms are correlated or not is an interesting 
and yet to be determined issue. Given the large top quark mass 
($m_t =175.6\pm 5.5$~GeV \cite{topmass}), as high as the EWSB scale,
it has been speculated that the top quark may play a special role
for the EWSB.  One of such ideas is that
some new strong dynamics may involve a composite Higgs
sector to generate the EWSB and
to provide a dynamical origin for the top quark 
mass generation ({\it e.g.,} 
the top-condensate/top-color models \cite{topCrev}).
Another idea is realized in the supersymmetric
theories in which the EWSB is driven radiatively by the large top quark
Yukawa coupling \cite{SUSY-rad}.

In the minimal standard model (SM) there is only 
one Higgs doublet, which 
leaves a physical neutral scalar boson 
as the remnant of the spontaneous EWSB.
The Yukawa couplings of the SM Higgs are determined from 
the relevant SM fermion masses divided by the vacuum expectation
value (VEV), $v\simeq 246$~GeV. 
Thus, aside from the coupling to the heavy top 
quark, all the other SM Yukawa couplings are highly suppressed,
independent of the Higgs boson mass.
For the top-condensate/topcolor type of models \cite{topCrev}, with a
composite Higgs sector, the new strong dynamics associated with 
the top sector
plays a crucial role for generating the large top mass and (possibly) 
the $W,Z$ boson
masses. As to be discussed below, in this scenario,
the interactions of the top-bottom sector with the 
composite Higgs bosons are different from that in the SM.
Due to the infrared quasi-fixed-point structure \cite{IRQFP}
and the particular boundary conditions at the compositeness scale, the
bottom Yukawa coupling to the relevant Higgs scalar is naturally 
as large as that of the top in such models. 
In the minimal supersymmetric extension of the SM (MSSM) 
\cite{mssm}, there are two Higgs doublets, whose mass
spectrum includes two neutral scalars ($h$ and $H$), 
a pseudoscalar ($A$) and a pair of charged scalars $(H^\pm)$.
The MSSM Higgs sector contains two free parameters 
which are traditionally chosen
as the ratio of the two Higgs VEV's ($\tan\beta = v_u/v_d$) 
and the pseudoscalar mass ($m_A$).
A distinct feature of the MSSM is that in the large $\tan\beta$ region
the Higgs Yukawa couplings to all the down-type fermions are 
enhanced by either $\tan\beta$ or $1/\cos\beta$.  
Among the down-type fermions, the bottom quark has the largest mass and so
the largest Yukawa coupling. Thus, it represents a likely place where new 
physics could reveal itself experimentally.
This common feature, the large bottom Yukawa coupling relative
to that of the SM, present in both types of the (conceptually 
quite distinct) models discussed above, serves as the theoretical
motivation for our analysis.

Since the third family $b$ quark, as the weak isospin
partner of the top, can  have large 
Yukawa coupling with the Higgs 
scalar(s) in both composite and supersymmetric models, 
we recently proposed \cite{bbbb_prl}
to use the $b$ quark as a probe of possible non-standard
dynamics in Higgs and top sectors.
In fact, because of the light $b$ mass ($\simeq 5$~GeV)
relative to that of the top ($\simeq 175$~GeV),
the production of Higgs boson associated with $b$ quarks
($\ppbar /pp \to \phibb \to \bbbb$) may be experimentally
accessible at the Fermilab Tevatron\footnote{
A $p \bar{p}$ collider with $\sqrt{s} = 2$\, TeV.} 
or the CERN Large Hadron Collider (LHC)\footnote{
A $p p$ collider with $\sqrt{s} = 14$\, TeV.}, even though the large
top mass could render associated Higgs production
with top quarks ($\ppbar /pp \to \phi t \bar{t}$) infeasible.
As we will show, this makes it possible for the Run~II of the Tevatron
and the LHC to confirm the various models in which the $b$-quark
Yukawa coupling is naturally enhanced relative to the SM prediction.
However, if the $\phi b\bar{b}$ signal is not found,
the Tevatron and the LHC can put stringent constraints 
on the models with either a composite or a supersymmetric Higgs 
sector, in which the Yukawa coupling of the Higgs boson(s) and bottom
quark is expected to be large.
In \cite{dai}, this reaction was explored at
the LHC and the Tevatron\footnote{
In \cite{dai}, a $1.8$~TeV $p\bar{p}$ Tevatron was assumed with an
integrated luminosity of 30~fb$^{-1}$, and
the squark mixings of the MSSM were neglected.} 
to probe the supersymmetry (SUSY) parameters of the MSSM.  
The conclusion
was that, with efficient $b$-tagging, useful information concerning
the MSSM could be extracted from either the LHC or a high luminosity
Tevatron.  In this work, we expand upon earlier results
\cite{bbbb_prl} in which it was concluded that even the
much lower integrated luminosity of the Tevatron Run~II can provide 
useful information through this reaction, provided an optimized search
strategy is employed.  We begin with a
model-independent analysis in Sec.~II, 
considering relevant backgrounds and determining an effective
search strategy to extract a signal from the backgrounds.
We then apply these results to constrain
both composite and supersymmetric models in Secs.~III and ~IV.
In Sec.~III, we first analyze the constraints on the two 
Higgs doublet extension of 
top-condensate model \cite{tt-2HDM} and the 
topcolor model 
\cite{topcolor}, and then remark upon the recent dynamical 
left-right model \cite{tt-lindner} (as a natural 
extension of the minimal 
top-condensate model \cite{BHL})
which also generically predict a large bottom Yukawa coupling.
In Sec.~IV, after deriving the exclusion contours on the $m_A$-$\tanb$
plane of the MSSM, we further analyze its implication on the 
supergravity\cite{susyrev} and gauge-mediated\cite{GMSB} models
of soft SUSY breaking that naturally predict a large $\tan\beta$.
The comparison of our Tevatron Run~II results with the LEP~II bounds 
(from the $Zh$ and $hA$ channels) is also presented,
illustrating the complementarity of our analysis to other 
existing Higgs search strategies. Final conclusions are given in
Sec.~V.

\section{Signal and Background}
\indent \indent

We are interested in studying production of $\phibb \to \bbbb$
at the Run II of the Tevatron and the LHC.  
The signal events result from QCD production of a 
primary $b \bar{b}$ pair, with a Higgs boson ($\phi$) radiated
from one of the bottom quark lines (c.f. Fig. \ref{sigfeynfig}).
The Higgs boson then decays into
a secondary $b \bar{b}$ pair to form a $\bbbb$ final state.
Because our detection strategy relies upon observing the primary $b$
quarks in the final state (and thus demands that they have large
transverse momentum), our calculation 
of the $\phibb$ signal rate from diagrams such as those
shown in Fig. \ref{sigfeynfig} is expected to be reliable.  This is in
contrast to the {\it inclusive} rate of $\phi$ production 
at a hadron collider, in which one does not require a final
state topology with four distinct jets.  In this case a calculation
based upon Feynman diagrams such as those shown in 
Fig.~\ref{sigfeynfig} may not be reliable.
It would be better to consider the Higgs boson production via 
bottom quark fusion, such as
$b \bar{b} \to \phi$ and $g b \to \phi b$, with cares to avoid
double counting its production rate \cite{dicus}.
(This calculation would resum some large logarithms which are included 
in the definition of the 
bottom parton distribution function within the proton.)
 We have chosen to search in the four jet
final topology because the QCD background for 3 jets is much
larger than that for 4 jets, and thus it would be more
difficult to extract a 3 jet signal.
Since the signal consists of four $b$ (including $\bar b$) jets,
the dominant backgrounds at a hadron collider come from 
production of $\Zbb \to \bbbb$ (c.f. Fig. \ref{zbbfeynfig}),
purely QCD production of $\bbbb$ (c.f. Fig. \ref{bbbbfeynfig})
\cite{bbbbbackground} and
$\bbjj$, where $j$ indicates a light quark or a gluon (c.f.
Fig. \ref{bbjjfeynfig}) which can occasionally produce a $b$-jet like
signature in the detector. 

In order to derive model-independent bounds on 
the couplings of the scalar particles with
the bottom quark, we consider $K$, the square-root
of the enhancement factor for the production of 
$\phibb \to \bbbb$ over the SM prediction.
By definition,
\be
K = \dis \frac{y_b}{(y_b)_{\rm SM}}  ~,
\ee
in which $(y_b)_{\rm SM} = \sqrt{2} \, m_b / v$ 
is the SM bottom Yukawa coupling 
and $y_b$ is the bottom Yukawa coupling in the new physics model under
the consideration\footnote{For simplicity, we ignore running effects in
$(y_b)_{\rm SM}$ in this section, treating the bottom quark mass as
5 GeV at all scales.  We will comment on the running effects of the
Yukawa coupling in the context of specific models of new physics in
Secs.~III and IV.}. 
The decay branching ratio of $\phi$ to $b\bar b$ is model-dependent, and
is not included in the calculations of this section. 
(Namely, the decay branching ratio Br$(\phi\to b\bar b)$ 
is set to be one).
We will properly take it into account for the specific models to be 
discussed in the following sections.

We compute the signal and the backgrounds at the parton level, using 
leading order (LO)
results from the MADGRAPH package \cite{madgraph}
for the signal and the backgrounds,
including the sub-processes initiated by $q \bar q$ and $g g$
(and in the case of $\bbjj$, $q g$ and $\bar q g$).
While the complete next-to-leading order (NLO) calculations are 
not currently
available for the signal or background cross sections, we draw upon
existing results for high $p_T$ 
$b \bar b$ production at hadron colliders \cite{kfac} and
thus estimate the NLO effects by including a
$k$-factor of 2 for all of the signal and background
rates.  (NLO effects to the $p p \to \phi t \bar{t}$ cross section
in the limit $s \gg m^2_t \gg m^2_\phi$ were explored in
\cite{dawsonreina}. It was found that 
QCD $k$-factor is on the order of 1.5,
but this limit is not expected to provide a very good estimate to the
$\phibb$ rate at the Tevatron since the corresponding condition 
$m_b^2\gg m_\phi^2$ no longer holds. 
Furthermore, because the mass of the bottom 
quark is much less than that 
of the top quark, we expect that the $k$-factor for the $\phibb$ 
production rate to be larger than that for the $\phi t\bar t$ rate.)
In the end of this section, we will also estimate 
the uncertainty in the calculation of the 
signal and background cross sections
based upon the above prescription.
We use the CTEQ4L \cite{cteq} parton distribution functions
(PDFs) and set the factorization scale, $\mu_0$, to the average of
the transverse masses of the primary $b$ quarks, and the boson 
($\phi$ or $Z$) transverse mass\footnote{
The transverse mass of particle $i$ is given by 
$m^{(i)}_{T} \equiv \sqrt{ m_i^2 + {p^{(i)}_T}^2}$.}
for the $\phibb$ and $\Zbb$ processes, and use a
factorization scale of $\mu_0 = \sqrt{\hat s}$,
where $\hat s$ is the square of the partonic center of mass energy,
for the $\bbbb$ and $\bbjj$ background processes.
It is expected that a large part of the total
QCD $\bbbb$ and $\bbjj$ rates
at the Tevatron or LHC energies will come from fragmentation effects,
which we have neglected in our matrix element calculation.  
However, due to the
strong $p_{T}$ and isolation cuts 
which we will impose for improving the
signal-to-background ratio (explained below),
we expect that these effects will be
suppressed, and thus will only have a small effect on our results.
Similarly, we expect that after imposing the necessary kinematic cuts,
the signal and the background rates are less 
sensitive to the above choice 
of the factorization scale.
In this section, unless otherwise noted,
we will restrict our discussion of numerical results
to a signal rate corresponding to a scalar mass of 
$m_{\phi} = 100$ GeV, and an enhancement factor of 
$K = m_t / m_b \approx 40$.  
We will consider the experimental limits which may be placed on $K$ as
a function $m_{\phi}$ below.

In order to simulate the detector acceptance, we require the $p_{T}$
of all four of the final state jets to be  $p_{T} \geq 15$\,
GeV, and that they lie in the central region of the detector,
with rapidity $|\eta | \leq 2$.  
We also demand that the jets are resolvable as
separate objects, requiring a cone separation 
of $\Delta R \geq 0.4$, where 
$\Delta R \equiv \sqrt{ {\Delta \varphi}^{2} + {\Delta \eta}^{2} }$.
($\Delta \varphi$ is the separation in the azimuthal angles.)
In the second column of Table~\ref{cutstab} we present the number of
events in the signal and background processes at the Tevatron Run~II
which satisfy these
acceptance cuts, assuming 2 \ifb of integrated luminosity.
As can be seen, the large background makes it
difficult to observe a signal in the absence of a carefully
tuned search strategy to enhance the signal-to-background ratio.
In presenting these numbers, we have assumed that it will 
be possible to trigger on events containing high $p_{T}$ jets 
(and thus retain all of the signal and background events).
This capability is essential for our analysis.

The typical topology of the bottom quarks in the
signal events is a ``lop-sided'' structure in which one of the
bottom quarks from the Higgs decay has a rather high $p_{T}$ of
about $m_{\phi} / 2$, whereas the other three are typically much softer.
Thus, the signal
events typically have one bottom quark which is much more energetic
than the other three.  On the other hand, 
the QCD $\bbbb$ (or $\bbjj$) background is typically much more 
symmetrical, with pairs of bottom
quarks (or fake $b$'s) with comparable $p_T$.
In order to exploit this, we order the
$b$ quarks by their transverse momentum, 
\be
p^{(1)}_{T} \geq p^{(2)}_{T} \geq p^{(3)}_{T} \geq p^{(4)}_{T},
\ee
and require that the bottom quark with highest transverse momentum
have $p^{(1)}_{T} \geq 50$ GeV, and that $p^{(2)}_{T} \geq 30$ GeV
and $p^{(3,4)}_{T} \geq 20$ GeV.\footnote{
Here, we have corrected the values given 
in Ref.~\cite{bbbb_prl} due to a
Fortran error in evaluating the signal distribution. We thank 
S.~Mrenna for cooperation in checking the Monte Carlo
simulation and for his help in detecting this numerical error.} 
In Fig.~\ref{fig:pts},
we show the differential cross sections with
respect to $p^{(1)}_{T}$ after the acceptance cuts and 
with respect to $p^{(2)}_{T}$ after the $p^{(1)}_T$ cut
for the signal and the QCD $\bbbb$
background at the Tevatron Run II.  These plots
illustrate the advantage in isolating the signal from the background
provided by the asymmetric cuts on the final state $b$ quarks outlined
above.
In the third column of Table~\ref{cutstab} we show the effect
of these cuts on the signal and backgrounds.  As can be seen, these
cuts reduce the signal by about $60\%$, while drastically
reducing the QCD $\bbbb$ background by about $90\%$.

Since the $p_T$ spectrum of the leading jets is determined by the mass
of the scalar boson produced, the leading $p_T$ cuts can be optimized
to search for a particular $m_{\phi}$.  The results 
for several values of
$m_{\phi}$ are presented in Table~\ref{ptcutstab}.  As expected
from the discussion above, the optimal cut on $p^{(1)}_T$ is close to 
$m_{\phi} / 2$ whereas the optimal cut on $p^{(2)}_T$ is somewhat lower
(generally closer to $m_{\phi} / 3$).  
We adopt these optimized $p_T$ cuts
for each mass considered, when estimating the search reach of the
Tevatron or LHC.

Another effective method for reducing the QCD background is to tighten
the isolation cut on the bottom quarks.  
In the QCD $\bbbb$ background, one of the $b \bar b$ pairs is 
preferentially produced from gluon splitting. Because of the collinear
enhancement, the invariant mass of this $b\bar b$ 
pair tends to be small,
and the $\Delta R$ separation of these two $b$'s prefers to be as small as
possible. On the contrary, in the signal events, the invariant mass of
the $b\bar b$ pair from
the $\phi$-decay is on the order of $m_\phi$, and the
$\Delta R$ separation is large because the angular distribution of $b$
in the rest frame of the scalar $\phi$ is flat.  Thus, by increasing
the cut on $\Delta R$ to $\Delta R \geq 0.9$ we can improve the
significance of the signal.  As shown in column four of 
Table~\ref{cutstab}, this cut further decreases the signal 
by about $30\%$, and the QCD $\bbbb$ 
background by about $65\%$. 
In the end, their event rates are about the same.

One can further improve the significance of the signal by attempting to
reconstruct the mass of the scalar resonance.  This can be difficult in
principle, because one does not know {\it a priori} what this mass is,
or which bottom quarks resulted from the $\phi$ decay in a given event.
It may be possible to locate the peak in the invariant mass distribution
of the secondary $b$ quarks resulting from the $\phi$ decay, though with
limited statistics and a poor mass resolution this may prove
impractical.  However, one can also scan through a set of masses, and
provide 95\% C.L. limits on the presence of a Higgs boson 
(with a given enhancement to the cross section, $K$) in the $\bbbb$
data sample for each value of $m_{\phi}$ in the set.
In order to do this, we assume a Higgs mass, and find the
pair of $b$ quarks with invariant mass which best reconstructs this
assumed mass.
We reject the event if this ``best reconstructed'' mass
is more than $2 \Delta m_{\phi}$ away from our assumed mass, where
$2 \Delta m_{\phi}$ is the maximum of either twice
the natural width of the scalar
under study ($\Gamma_{\phi}$) or the twice 
experimental mass resolution\footnote{
We estimate the experimental mass resolution for 
an object of mass $m_{\phi}$
to be $\Delta m_{\phi} = 0.13 \, m_{\phi} \, 
\sqrt{ 100 \,{\rm GeV} / \, m_{\phi}}$.
Under this assumption, the natural width of the bosons in the
specific models of new physics considered in Secs. III and IV is
usually smaller than this experimental mass resolution.}.
As shown in the fifth column of
Table~\ref{cutstab}, this cut has virtually no effect on the signal or
$\Zbb$ background (for a 100 GeV Higgs)
while removing about another $10\%$ of the $\bbbb$ background.

As will be discussed below, the natural width of
the Higgs bosons in both the MSSM and the models of strong EWSB that we
wish to probe in this paper
are generally much smaller than our estimated
experimental mass resolution, and thus one might think that an improved
experimental mass resolution could considerably improve the limits
one may place on a scalar particle with a strong $b$ interaction.
However,  the models in which we are interested generally have
one or more nearly mass-degenerate bosons with similarly enhanced bottom
Yukawa couplings.  If the extra scalars are much closer in mass than
the experimental mass resolution
(and the natural width of the bosons), the signal can thus include
separate signals from more than one of them.  Thus there is potentially
a trade-off in the $\Delta M$ cut (cf. Table I) between 
reduction of the background
and acceptance of the signal from more than one scalar resonance.  In
order to estimate the potential improvement for discovering a single
Higgs boson, we have examined the effect on the significance one obtains
if the cut on the invariant mass which best reconstructs $m_{\phi}$ is
reduced to $\Delta m_{\phi}$ as opposed to 
$2 \Delta m_{\phi}$ as was considered
above.  We find that this improved mass resolution further reduces the
QCD $\bbbb$ background by about another $40\%$.  Assuming four $b$
tags (as discussed below), this improved mass resolution increases the
significance of the signal from about 12.2 to 14.6, which will
improve the model-independent lower bound on $K$ by about $10\%$.
Thus, an improved mass resolution would most likely be helpful in
this analysis.

Another method to further suppress background rate is to observe that
in the background events,  the
$b$ quarks whose invariant mass best reconstructs $m_{\phi}$ 
come from the
same gluon.  This is because, after imposing all the kinematical cuts 
discussed above (cf. Table~I), the matrix
elements are dominated by Feynman diagrams in which one very far
off-shell gluon decays into a $b \bar{b}$ pair, as opposed to
interference of many production diagrams, which dominates the lower
invariant mass region.  
Thus, for $m_{\phi}$ greater than about 100 GeV,
the background event produces $b$ quarks with the characteristic
angular distribution of a vector decaying into fermions,
$1 + \cos^2 \theta$, in the rest frame of the $b \bar{b}$ system.
This is distinct from the signal distribution, which comes from
a scalar decay, and is flat in $\cos \theta$.  Thus, for masses above
$100$ GeV, we further require $|\cos \theta| \leq 0.7$
after boosting back to the rest frame of the $b \bar{b}$ 
pair which we have identified as coming from the scalar boson $\phi$.

In order to deal with the large QCD $\bbjj$ background, it is important
to be able to distinguish jets initiated by $b$ quarks from those
resulting from light quarks or gluons. We estimate the probability to
successfully identify a $b$ quark passing the acceptance cuts outlined
above to be 60\%, with a probability of 0.5\% to misidentify a jet
coming from a light quark or gluon as a $b$ jet \cite{tev2000}.
In Table~\ref{btagtab} we show the resulting number of 
signal and background events passing our optimized cuts at the
Tevatron, assuming 2 \ifb of integrated luminosity, after
demanding that two or more, three or more, or 
four $b$-tags be present in the
events, and the resulting significance of the signal (computed as
the number of signal events divided by the square root of the number
of background events).  We find that requiring 3 or more 
$b$-tags results in about the same significance of
$12.2 \sigma$ as requiring 4
$b$-tags.  However,
we see that for the chosen parameters ($m_{\phi} = 100$ \, GeV and 
$K = m_t / m_b \approx 40$), even with only 2 or more $b$-tags, one
arrives at a significance of about $3 \sigma$, and thus has some ability
to probe a limited region of parameters.
From the large significance, we see that the 
Tevatron may be used to place
strong constraints on Higgs particles with enhanced bottom quark Yukawa
couplings, and that the ability to tag 3 or more of the bottom quarks
present in the signal can probe a larger class of models (or parameter
space of the models)
as compared to what is
possible if only 2 or more of the bottom quarks are tagged.
In the analysis below, to allow for the possibility that the $\bbjj$
background may be somewhat larger than our estimates, we
require 4 $b$-tags, though as we have demonstrated above, we do not
expect a large change in the results if 3 or 4 $b$-tags were
required instead.

This analysis can be repeated for any value of $m_\phi$, using
the corresponding $p_T$ cuts shown in Table \ref{ptcutstab}.
It is interesting to note that the signal composition in terms of the
$g g$ or $q \bar{q}$ initial state depends on the collider type and 
the mass of the produced boson, which controls the
type of PDF and the typical region of $x \sim m^2_{\phi} / S$ 
at which it is evaluated.  
At the Tevatron, for
$m_\phi = 100\,$ GeV, the signal is 99\% $g g$ initial state before
cuts, and 87\% after cuts, while for $m_\phi = 200\,$ GeV,
it is  99\% $g g$ initial state before cuts, and 85\% after cuts.
Thus, at the Tevatron, one ignores about 15\% of the signal if one
relies on a calculation employing only the $g g$ initial state.
At the LHC, for $m_\phi = 100$, the signal is 
very close to 100\% $g g$ initial state
before cuts and 99\% after cuts, and for $m_\phi = 500\,$ GeV,
it is 99\% $g g$ initial state before cuts, and 99\% after cuts.
This indicates that at the LHC, very accurate results are possible from a
calculation considering only the $g g$ initital state.
The resulting numbers of signal and (total) background events after
cuts for various boson masses are shown in Table~\ref{eventnumtab}.

From these results, one may derive the minimum value of $K$, $K_{\min}$,
for a scalar boson with mass $m_\phi$ to 
be discovered at the Tevatron or
the LHC via the production mode $b\bar{b}\phi (\to b\bar{b})$.
Similarly, if signal is not found, one can exclude models which 
predict the enhancement factor $K$ to be larger than $K_{\min}$.
To give a model-independent result, we
assume that the width of the $\phi$ is much less 
than the estimated experimental
mass resolution defined above, which is the case for the models studied
in this paper.  We determine $K_{\min}$ by noting that in
the presence of a Higgs boson with enhanced bottom Yukawa couplings, the
number of expected signal events passing our selection criterion
is given by $N_S = K^2 \, N_S^{(SM)}$,
where $N_S^{(SM)}$ is the number of signal events expected for a scalar
of mass $m_{\phi}$ with SM coupling to the $b$ quark [assuming
Br$(\phi\to b\bar{b})=1$], whereas the number of
background events expected to pass our cuts, 
$N_B$, is independent of $K$.
Thus, requiring that
no $95\%$ C.L. deviation is observed in the $\bbbb$ data sample
(and assuming Gaussian statistics) determines 
\be
K_{\min} = \sqrt{ \frac{1.96 \, \sqrt{N_B} }{N_S^{(SM)} } },
\label{kmindef}
\ee
where $1.96 \sigma$ is the $95\%$ C.L. in Gaussian statistics.
In Fig.~\ref{kminfig}, we show the
resulting 95\% C.L. limits one may impose 
on $K_{\min}$ as a function of 
$m_{\phi}$ from the Tevatron with 2, 10, 
and 30 \ifb and from the LHC with
100 \ifb, as well as the discovery reach of the LHC at the 
$5 \sigma$ level.
Our conclusions concerning the LHC's ability to probe a Higgs
boson with an enhanced $b$ Yukawa coupling are very similar to those
drawn in \cite{dai}, but are considerably more
optimistic than those in \cite{froidevaux}, where the conclusion was
that the $\bbjj$ background is considerably larger than our estimate
(though there are elements of the search strategy which differ between
those of \cite{froidevaux} and ours as well, and their simulation of the
ATLAS detector is certainly more sophisticated).
In \cite{froidevaux} the backgrounds were simulated 
using PYTHIA \cite{pythia}
to generate two to two hard scatterings and then generating the
additional jets from a parton showering algorithm.  As noted above, in
the light of the strong (ordering of) $p_T$ 
and isolation cuts applied to select the
signal events, we feel that a genuine four body matrix element
calculation such as was used in our analysis provides a more reliable
estimate of this background.

We have examined the scale and PDF dependence of our calculation for the
signal and background rates at the Tevatron, 
and find that in varying the scale between one half and twice its
default choice (defined above),
$\mu = \mu_0 / 2$ and $\mu = 2 \mu_0$, the $\phibb$
signal and $\Zbb$ background rates both vary from the result at
$\mu = \mu_0$ by about $30\%$, while the
$\bbbb$ and $\bbjj$ backgrounds vary by about $45\%$.  This strong scale
dependence is indicative of the possibility of large higher order 
corrections to the leading order rate.  Thus, in order to better
understand the true signal and background rates, it would be useful to
pursue these calculations to NLO.  We have also compared the difference
in the results from the MRRS(R1) PDF \cite{mrrs} and the CTEQ4L 
PDF, and find a variation of about $10\%$ in the resulting 
signal and background rates.  Since these separate sources of
uncertainty (from PDF and scale dependence)
are non-Gaussianly distributed, there is no way to rigorously
combine them.  Thus, we 
conservatively choose to add them linearly, finding a total 
uncertainty of about $40\%$ in the signal rate ($N^{(SM)}_S$),
and $50\%$ in the background rate ($N_B$).  
From the derivation of $K_{\min}$ above, we see
that these uncertainties in signal and background rate 
(which we assume to be uncorrelated)
combine to give a fractional uncertainty in
$K_{\min}$,
\be
\frac{\delta K_{\min}}{K_{\min}} = \sqrt{ 
{\left( \frac{\delta N_S^{(SM)}}{2 \, N_S^{(SM)}}\right)}^2 + 
{\left( \frac{\delta N_B}{4 \, N_B}\right)}^2},
\label{kminerr}
\ee
where $\delta N_S^{(SM)}$ and $\delta N_B$ are the absolute
uncertainties in $N_S^{(SM)}$ and $N_B$, respectively.  From this
result, we see that in terms of a more precise theoretical
determination of $K_{\min}$, one gains much
more from a better understanding
of the signal rate than a better determination of the backgrounds.
Applying our estimation of the uncertainty from PDF and
scale dependence to Eq.~(\ref{kminerr}),
we find an over-all theoretical uncertainty in $K_{\min}$ of about
$25\%$.

\section{CONSTRAINTS AND IMPLICATIONS ON 
DYNAMICAL MODELS WITH STRONGLY COUPLED ELECTROWEAK SECTOR }

The observed large top mass, of the order of the electroweak scale,
singles out top quark from all the other light fermions. This makes
the top-condensate/topcolor scenario particularly 
attractive \cite{topCrev}.  
In this section, we analyze the strongly interacting scenario 
of the EWSB with a composite Higgs sector. 
We consider top-condensate/topcolor type of models 
\cite{tt-2HDM,topcolor,tt-lindner,BHL,topcolor2,topcolor3,Nambu,top-seesaw}
in which new strong dynamics associated with the top quark sector
plays a crucial role in the generation of the 
top quark and the $W,Z$ boson masses.
As we have emphasized, since the bottom quark is
the weak isospin partner of the top quark, its 
interaction to the Higgs sector can be 
closely related to that of the top quark. In the
top-condensate/topcolor scenario to be analyzed below, 
the $b$ quark Yukawa
coupling (to the relevant scalar) 
is {\it naturally large,} of the same order as
the top Yukawa coupling [$\sim O(1)$],  due to the quasi-infrared 
fixed point structure\cite{IRQFP} and the proper boundary conditions
at the compositeness scale.  This can give distinct experimental
signatures at the Tevatron and the LHC. In the following, we shall
analyze two specific models in this scenario, and derive 
the constraints (or discovery reach) expected at 
the Tevatron and the LHC based upon the 
model-independent results  
in Sec.~II. Finally, we shall analyze the 
dynamical left-right
symmetric extension \cite{tt-lindner} 
of the minimal top-condensate model \cite{BHL}.

\subsection{
In Two Higgs Doublet Extension of the Minimal Top-Condensate Model}

Since the minimal top-condensate model 
(with three families) \cite{BHL} was ruled out due to
predicting a too large top mass ($\sim 220-250$~GeV)
to reconcile with experimental data,
we consider the minimal two Higgs doublet
extension (2HDE) of the top-condensate model proposed in 
Ref.~\cite{tt-2HDM}, which predicts a smaller value of $m_t$. 
Though the simplest 2HDE of the top-condensate model
may not provide enough reduction of the $m_t$ value to match the
Tevatron measurement, it is not out
of question to incorporate some further
improvements \cite{topCrev}, 
{\it e.g.} including the recently proposed seesaw-type
top-condensation  \cite{top-seesaw}, to achieve a realistic $m_t$. 
In the supersymmetrized
version of the top-condensate model \cite{susy-topc},
it is possible to derive the correct
top mass while keeping the similar boundary 
conditions and the quasi-infrared 
fixed point structure which ensures the large $b$ quark (and also 
$\tau$ lepton) Yukawa
couplings.  In this subsection we analyze the simplest 
2HDE of the top-condensate model constructed in 
Ref.~\cite{tt-2HDM},
for illustration. 

The starting point of the model is to consider the SM without 
an elementary Higgs boson,
but with Nambu--Jona-Lasinio (NJL) type of four-Fermi 
interactions \cite{NJL} generated at the cut-off scale $\Lambda$,
where the new physics takes place. 
For the third generation quarks, the $SU(2)_L\otimes U(1)_Y$ invariant
4-Fermi couplings can be written as~\cite{tt-2HDM}:
\be
{\cal L}_{4F} = 
G_t\left(\Psibar_Lt_R\right)\left(\tbar_R\Psi_L\right) + 
                G_b\left(\Psibar_Lb_R\right)\left(\bbar_R\Psi_L\right) +
                G_{tb}\left[\left(\Psibar_Lb_R\right)
                            \left(\tbar^c_R\widetilde{\Psi}_L\right) 
                + {\rm h.c}\right] \, ,
\label{eq:NJL}
\ee
where the summation over color indices is 
implied in the round parentheses
and $~\widetilde{\Psi}=(-b^c, t^c)^T~$. Then, just below
 the scale $\Lambda$, two
composite Higgs doublets, $\Phi_t$ and $\Phi_b$, 
can be introduced via the
auxiliary field method \cite{auxiliary} with the interaction Lagrangian
\be
{\cal L}_{SF} =
-\mu_t^2\Phi_t^{\dag}
\Phi_t+\left(\Psibar_L{\Phi}_tt_R+{\rm h.c.}\right)
-\mu_b^2\Phi_b^{\dag}
\Phi_b+\left(\Psibar_L{\Phi}_bb_R+{\rm h.c.}\right)
-\mu_{tb}^2\left(\Phi_t^{\dag}
\Phi_b+{\rm h.c.}\right)  .
\label{eq:Yukawa0}
\ee
To diagonize the $\Phi_t$ and $\Phi_b$ mass terms, one needs to 
introduce the mixing angle $\alpha$ defined by:
$~~
\dis\alpha = 2\mu_{tb}^2/(\mu_t^2-\mu_b^2)
~$.~
In Eqs.~(\ref{eq:NJL}) and (\ref{eq:Yukawa0}), 
the mixing term proportional to $G_{tb}$ 
or $\mu_{tb}$ is important to break the Peccei-Quinn $U(1)$ symmetry and
to generate a nonzero mass for the pseudoscalar.
The low energy Lagrangian at the scale $\mu (<\Lambda )$ 
can be deduced from
(\ref{eq:Yukawa0}) via the renormalization group (RG) 
evolution \cite{RGE} which
defines the effective low energy couplings. Thus, at the scale $\mu$~,
\be
\begin{array}{ll}
{\cal L}^{(r)}_{SF} = &
Z_{\Phi_t}^{\f{1}{2}}y_t\left(\Psibar_L{\Phi}_tt_R+{\rm h.c.}\right)+
Z_{\Phi_b}^{\f{1}{2}}y_b\left(\Psibar_L{\Phi}_bb_R+{\rm h.c.}\right)+
\\
& Z_{\Phi_t}(D^\mu\Phi_t)^{\dag}(D_\mu\Phi_t) +
Z_{\Phi_b}(D^\mu\Phi_b)^{\dag}(D_\mu\Phi_b) +
V(Z_{\Phi_t}^{\f{1}{2}}\Phi_t , Z_{\Phi_b}^{\f{1}{2}}\Phi_b) ~,
\end{array}
\label{eq:YukawaR}
\ee
where $~V~$ is the renormalized Higgs potential 
for $\Phi_t$ and $\Phi_b$.
As $\mu$ approaches to $\Lambda$, 
(\ref{eq:YukawaR}) should match with the bare Lagrangian
(\ref{eq:Yukawa0}) to result in the proper 
boundary conditions to be used in the renormalization group
analysis \cite{tt-2HDM}.
It turns out that the compositeness of both $\Phi_t$ and $\Phi_b$
can be achieved only for the boundary condition
$~y_t(\Lambda )=y_b(\Lambda )\equiv y_0\gg 1~$ 
\cite{tt-2HDM,topCrev}. Hence, 
\be
y_t(\mu )\approx y_b(\mu)~,~~~~~~~({\rm for~any}~ \mu <\Lambda ) ~.
\label{eq:gt=gb}
\ee
This is due to the fact that $y_t$ and $y_b$ 
satisfy similar RG equations
except for 
a small difference (in the $g_1^2$ term) originating from the different 
hyper-charges of the  $t$ and $b$ quarks \cite{RGE,topCrev}:
{\small 
\be
\begin{array}{l}
\dis\f{dy_t(\mu )}{d\ln\mu} =\f{1}{16\pi^2}
\left[\left(\f{3}{2}+N_c\right)y_t^2(\mu )+\f{1}{2}y_b^2(\mu )
-3\left(N_c-\f{1}{N_c}\right)g_3^2(\mu )-\f{9}{4}g_2^2(\mu )
-\f{17}{12}g_1^2(\mu )\right]y_t(\mu ),\\[0.6cm]
\dis\f{dy_b(\mu )}{d\ln\mu} =\f{1}{16\pi^2}
\left[\left(\f{3}{2}+N_c\right)y_b^2(\mu )+\f{1}{2}y_t^2(\mu )
-3\left(N_c-\f{1}{N_c}\right)g_3^2(\mu )-\f{9}{4}g_2^2(\mu )
-\f{5}{12}g_1^2(\mu )\right]y_b(\mu ),\\[0.3cm]
\end{array}
\label{eq:RGE}
\ee
}
where $N_c=3$ for the QCD theory, and 
$\mu \geq m_t\sim M_{\rm Higgs}$. 
Because of the infrared quasi-fixed point structure 
\cite{IRQFP,RGE} of the two Higgs doublet model, the relation
(\ref{eq:gt=gb}) holds well as long as 
$y_b(\Lambda ), y_t(\Lambda ) \geq 1$. 
This is true even for the case where
$y_b$ is chosen to be significantly lower than $y_t$
at the compositeness scale $\Lambda$.
This running behavior is shown in
Fig.~\ref{Fig:TC-RG}, which confirms the large value of $y_b$
at the weak scale. We have 
also examined possible threshold effects due to 
different values of the Higgs masses and found the above 
conclusion unchanged.

The two composite Higgs doublets develop VEVs from the 
condensation of
$<\hspace*{-0.15cm}\tbar t\hspace*{-0.15cm}>$ and 
$<\hspace*{-0.15cm}\bbar b\hspace*{-0.15cm}>$
 (determined by the gap equations),
so that
$~~
<\hspace*{-0.15cm}\Phi_t\hspace*{-0.15cm}> = 
\left(v_t, 0\right)^T/\sqrt{2}~$ and
$~<\hspace*{-0.15cm}\Phi_b\hspace*{-0.15cm}> = 
\left(0, v_b\right)^T/\sqrt{2}~$.~
Since the masses of the $t$ and $b$ quarks are given by
$~m_{t,b}(\mu )=y_{t,b}(\mu )v_{t,b}(\mu )/\sqrt{2} ~$,
and the Yukawa couplings $~y_b$ and  $y_t$ are about the same, of the 
$O(1)~$ at the weak scale
[cf. (\ref{eq:gt=gb}) and Fig.~\ref{Fig:TC-RG}], this model 
naturally {\it predicts} a large $\tan\beta$:
\be
\dis \tan\beta =\f{v_t(m_t)}{v_b(m_t)}
\approx \f{m_t(m_t)}{m_b(m_t)} \approx 55 \gg 1 ~.
\ee
Here,
$m_{b(t)}(m_t)$ is the running bottom (top) mass at the scale
$m_t$. The running values of $m_b(\mu )$ and $m_t(\mu )$ 
are derived\cite{Barger} from the measured physical pole
masses $m_b^{\rm pol}\simeq 5$~GeV \cite{PDG}
and $m_t^{\rm pol}\simeq 175$~GeV,
and are dominated by the QCD evolution at scales
$\lae O(M_{\rm Higgs})$.
At the one-loop level, the
relation between the pole quark mass $m_q^{\rm pol}$
and the $\overline{\rm MS}$ QCD running mass at the scale 
$\mu =m_q^{\rm pol}$ is:
\be
m_q(m_q^{\rm pol}) = 
m_q^{\rm pol}\left[1+
\dis\f{4\alpha_s(m_q^{\rm pol})}{3\pi}\right]^{-1}~.
\ee
When running upward to any scale $\mu$,
\be
m_q(\mu )=m_q(m_q^{\rm pol})\dis\f{c\left[\alpha_s(\mu )/\pi\right]}
                        {c\left[\alpha_s(m_q^{\rm pol})/\pi\right]}~,
\label{eq:running}
\ee
where $~c[x] = (23x/6)^{12/23}\left[1+1.175x\right]$ for 
$m_b^{\rm pol}<\mu <m_t^{\rm pol}$, and 
$~c(x) = (7x/2)^{4/7}\left[1+1.398x\right]$ for 
$\mu >m_t^{\rm pol}$ \cite{hdecay}.
Numerically,  $m_t(m_t^{\rm pol})\simeq 166$\,GeV and 
$m_b(m_t^{\rm pol})\simeq 3$\,GeV.

Analyzing the mass spectrum of the Higgs sector, we 
find that the lightest
scalar particle with the large Yukawa coupling to the bottom quark
is the pseudoscalar $P$ 
($=\sqrt{2}\left[\sin\beta {\rm Im}\Phi_b^0+\cos\beta 
{\rm Im}\Phi_t^0\right]
\sim \sqrt{2}{\rm Im}\Phi_b^0$)
with a mass
\be
M_P ~
\simeq \dis\f{v}{\sqrt{2}}|\lambda_4|^{1\over 2}
\left[ \f{\tan (\pi -2\beta )}{\tan 2\alpha}-1\right]^{-{1\over 2}}
~.
\label{eq:MP}
\ee
Since $~v=\sqrt{v_t^2+v_b^2}\simeq 246~$GeV, $M_P$ can be  
as low as $O(m_Z)$, depending on the
Higgs mixing angle $\alpha$ and the Higgs self-couplings 
$\lambda_4$ \cite{tt-2HDM}.
For instance, for $\Lambda =10^{15}$~GeV,
the mass $M_P$ is less than about $233$~GeV, and 
the decay branching ratio of $P$ to $b\bar b$ is about one.
Hence, this model predicts a light pseudoscalar that
couples to the bottom quark strongly through Yukawa interaction.

To discover or exclude this model at the Tevatron and the LHC
via measuring the production rate of four $b$ jets,
we need to make use of the model-independent results obtained in 
Sec.~II, Namely,
we need to compare the predicted
bottom Yukawa coupling [$y_b^P=y_b(\mu )\sin\beta$] 
of the pseudoscalar $P$ 
with the model-independent bound ($=K_{\min}y_{b0}^{\rm SM}$) derived
in Sec.~II, where the reference value $y_{b0}^{\rm SM}$ is arbitrarily
chosen to be $y_{b0}^{\rm SM}=\sqrt{2}m_b^{\rm pol}/v$ with 
$m_b^{\rm pol}\simeq 5$~GeV and $v\simeq 246$~GeV. 
This is equivalent to comparing  $y_b^P/y_{b0}^{\rm SM}$ with
$K_{\min}$, where $y_b^P$ is the running Yukawa coupling at the scale
$M_P$.  Note that at the weak scale the running effects of
the VEVs [mainly due to the electroweak corrections] are negligible
\cite{Barger}, and for $\mu\lae M_P$,
$m_b(\mu )=y_b(\mu )v_b=y_b^{\rm SM}(\mu )v$, so that
$y_b^P(\mu )=\sin\beta y_b(\mu )
=\tanb\sqrt{2}m_b(\mu )/v$. We can thus derive
a minimal $\tanb$ value for a given $M_P$ to discover such a model
at hadron colliders.
The result of $95\%$~C.L. exclusion contours is given in Fig.~8 for
various colliders. 
As shown in Fig.~\ref{Fig:TC-bound}, 
the 2~fb$^{-1}$ Tevatron Run~II data can exclude models with
$M_P$ up to $\sim 190$~GeV, if no signal is found. The entire
mass range of $P$ predicted in this model 
(less than 233\,GeV for $\Lambda =10^{15}$~GeV) 
can already be explored at the Tevatron 
Run~II with a 10~fb$^{-1}$ luminosity.

\subsection{In Topcolor Assisted Technicolor Model}

The minimal top-condensate model \cite{BHL} and its two-Higgs-doublet 
extension \cite{tt-2HDM} require fine-tuning 
the four-Fermi coupling(s) at the scale $\Lambda$ to be very close to
the critical value, but do not address the
dynamical origin of the effective coupling(s)
at the energy scale above $\Lambda$. 
The topcolor assisted technicolor models (TCATC) \cite{topcolor} 
were proposed to overcome such difficulties.
These models postulate a gauge structure
${\cal G}=
SU(3)_1\otimes SU(3)_2\otimes U(1)_1\otimes U(1)_2\otimes SU(2)_W$ 
at the scales above $\Lambda =O$(1~TeV)~.
The third family fermions couple
to $SU(3)_1\otimes U(1)_1$ gauge sector
with the same quantum numbers as those under the SM
QCD and $U(1)_Y$ interactions, 
while the first two family fermions couple to 
$SU(3)_2\otimes U(1)_2$ in a similar way. 
At the scale $\Lambda =O$(1~TeV),
$SU(3)_1\otimes U(1)_1$ is strong but not confining, and 
${\cal G}$ is spontaneously
broken down to 
${\cal G}_{\rm SM}=SU(3)_c\otimes U(1)_Y\otimes SU(2)_W$ 
due to an unspecified mechanism
which may or may not be related to the EWSB.
In consequence, massive gauge bosons of the 
color octet $B^a$ (colorons) and
the singlet $Z'$ are generated. Below this breaking scale
$\Lambda ={\min} (M_B,~M_{Z'})$, 4-Fermi interactions are
generated as follows:
\be
{\cal L}_{4F} =\dis \f{4\pi}{\Lambda^2}\left\{
\left[\kappa +\f{2\kappa_1}{9N_c}\right]
\left(\Psibar_Lt_R\right)\left(\tbar_R\Psi_L\right)+
\left[\kappa -\f{\kappa_1}{9N_c}\right]
\left(\Psibar_Lb_R\right)\left(\bbar_R\Psi_L\right) \right\}~.
\label{eq:LTC}
\ee  
After the fermions condensate, an effective Lagrangian with
two composite Higgs doublets ($\Phi_t$ and $\Phi_b$)
can be introduced as:
\be
{\cal L}_{SF} = 
  \left[ y_t\Psibar_L\Phi_tt_R +y_b\Psibar_L\Phi_bb_R +{\rm h.c}\right]
           -\Lambda^2 
\left[\Phi_t^{\dag}\Phi_t+\Phi_b^{\dag}\Phi_b\right],
\label{eq:LTCY}
\ee
with 
\be
y_t=\sqrt{4\pi (\kappa +2\kappa_1/9N_c)}~,~~~~~~
       y_b=\sqrt{4\pi (\kappa -\kappa_1/9N_c)}~.
\label{eq:yt-yb}
\ee
Here $\kappa$ and $\kappa_1$ originate from the strong $SU(3)_1$ and
$U(1)_1$ dynamics, respectively. The $U(1)_1$ force is attractive
in the 
$<\hspace*{-0.15cm}\tbar t\hspace*{-0.15cm}>$ channel but repulsive 
in the $<\hspace*{-0.15cm}\bbar b\hspace*{-0.15cm}>$ channel, such that
the top but not the bottom acquires dynamical mass under the condition
\be
y_b ~<~ y_{\rm crit}=\dis\sqrt{\f{8\pi^2}{3}} ~<~ y_t ~~.
\label{eq:crit-condition}
\ee
Equivalently, this implies that the composite Higgs $\Phi_t$ but not 
$\Phi_b$ develops a VEV, i.e., $v_t\neq 0$ and $v_b=0$, in contrast to
the simplest 2HDE of top-condensate model analyzed in the previous
subsection (where $v_{t,b}\neq 0$).  
For $\mu <\Lambda$, the composite Higgs doublets $\Phi_t$ and  
$\Phi_b$ develop the gauge invariant kinetic terms
\be
{\cal L}_{SF}^{\rm kin}= 
Z_{\Phi_t}(D^\mu\Phi_t)^{\dag}(D_\mu\Phi_t) +
Z_{\Phi_b}(D^\mu\Phi_b)^{\dag}(D_\mu\Phi_b) ~,~~~~
{\rm with}~~~~ Z_{\Phi_{t,b}}=
\dis\f{N_c}{8\pi^2}y_{t,b}^2\ln\f{\Lambda}{\mu}~,
\label{eq:kinetic}
\ee 
as well as a potential term ~$V(Z_{\Phi_t}^{\f{1}{2}}\Phi_t ,
 Z_{\Phi_b}^{\f{1}{2}}\Phi_b)$.~

From (\ref{eq:yt-yb}), we
note that unless $\kappa_1$ is unnaturally large  (compared 
to $\kappa$), both $y_t$ and $y_b$ should be close to the
critical value $~y_{\rm crit}=\sqrt{8\pi^2/3}\simeq 5.13~$ at the scale
$\Lambda$. At the lower energy scale $\mu (<\Lambda )$, $y_b$ is
still close to $y_t$, based on the RG analysis shown in 
Fig.~\ref{Fig:TC-RG}.
From the RG evolution, 
we find $~y_b(m_t)=2.7-2.1~$ and $~y_t(m_t)=2.9-2.2~$ for 
$\Lambda =1-10$~TeV, with the typical boundary conditions
$~y_t(\Lambda )=5.5~$ and $~y_b(\Lambda )=4.5~$. 
The precise boundary values
of $y_{t,b}(\Lambda )$ depends on the 
detailed dynamics of topcolor breaking
via the parameters $(\kappa ,~\kappa_1)$. But for $(\kappa ,~\kappa_1)$
not much larger than $O(1)$, $y_{b,t}(\Lambda )$ should be reasonably
close to the critical value $y_{\rm crit}$ at the scale $\Lambda$.
Furthermore, as shown in Fig.~\ref{Fig:TC-RG}, the infrared behavior of 
$y_{t,b}(\mu )$ at the weak scale is not sensitive to the possible
variations of their boundary values at the scale $\Lambda$. 
Therefore, this model generically predicts a large $y_b$ of $O(2-3)$ 
at the weak scale. 

Another essential feature of the TCATC model \cite{topcolor} 
is that the topcolor interaction must
not be responsible for the whole EWSB, but is mainly 
responsible for the top quark mass
generation. As a result, the dynamical scale can be as low as
$\Lambda =O(1)$~TeV (which avoids the severe fine-tuning needed in the
minimal models \cite{tt-2HDM,BHL})
and correspondingly, $v_t\simeq 64-97$~GeV (for $\Lambda =1-10$~TeV) 
according to the Pagels-Stokar formula \cite{PS}:
\be
\dis v_t^2 ~=~ \f{N_c}{8\pi^2}m_t^2\left(\ln\frac{\Lambda^2}{m_t^2}
+c_0\right)
\label{eq:f_pi}
\ee 
where $c_0 = O(1)$ is a constant.
The $b$ quark gets large portion of its mass 
from topcolor instanton effects \cite{topcolor}. 
The EWSB is mainly driven by the
usual extended technicolor (ETC) 
\cite{ETC0,ETC} (or the equivalent Higgs) interaction 
which gives small masses [$\lae O$(GeV)] 
to all fermions (including $t$ and $b$).
In addition, this model predicts three physical top-pions 
with masses around $O(m_t)$.
The smaller vacuum expectation value $v_t$
[estimated by (\ref{eq:f_pi}) in TCATC models], as compared to the
full VEV ($v \simeq 246$~GeV),
makes the Yukawa coupling of the top to $\Phi_t$ 
stronger than that in the SM,
which is consistent with the predictions from the RG analysis in
Fig.~\ref{Fig:TC-RG}. The large-$N_c$ calculation 
\cite{topcolor-more} suggests that
the neutral components $(h_b,~A_b)$ of $\Phi_b$ are degenerate
and can be the lightest Higgs bosons with masses of $O(100)$~GeV 
to a few hundred GeV.
Hence, this model also predicts light scalars which couple to 
the bottom quark strongly via Yukawa interaction.

To illustrate how Tevatron (Run~II) and LHC  
can test this model via the reaction
~$p\bar{p}/pp\to b\bbar h(\to b\bbar )+X$~,  we compare $y_b(\mu )$ 
(predicted in Fig.~\ref{Fig:TC-RG}) with the model-independent
bound on $K_{\min}y^{\rm SM}_{b0}$ derived in Sec.~II. Or, equivalently,
we compare $y_b(\mu )/y^{\rm SM}_{b}(\mu)$ with  
$K_{\min}y^{\rm SM}_{b0}/y^{\rm SM}_b(\mu )~
[=K_{\min}m_b^{\rm pol}/m_b(\mu )]$.
The result is shown in Fig.~\ref{Fig:TC-bound}b.
It is clear that 
the Tevatron Run~II can already provide useful information on this 
type of models. 
The LHC data can further extend the coverage of the mass range
up to 1\,TeV.
In Fig.~\ref{Fig:TC-LHCbound}, we present
the exclusion curve at the $95\%$~C.L. and the 
discovery reach at the 5$\sigma$ level
for the LHC with 100 \ifb of luminosity.
If we relax the mass degeneracy condition (suggested by the large-$N_c$ 
analysis) and assume that the degeneracy of $h_b$ 
and $A_b$ does not hold well enough to be within the mass 
resolution of the
detector (cf. Sec.~II), then the contours 
in  Figs.~\ref{Fig:TC-bound} and \ref{Fig:TC-LHCbound} 
will move up by an overall factor of $\sqrt{2}$
for most of the mass region. Fig.~\ref{Fig:TC-LHCbound} shows
that even in this non-degeneracy case, the LHC (with a 100~fb$^{-1}$
luminosity) can discover the Higgs boson  $h_b$ or $A_b$ with a mass
up to 1~TeV at the 5$\sigma$ level, since the theory curves always
lie above $y_b/y^{\rm SM}_b=100$ for 
$M_{h_b (A_b)}\geq 50$~GeV (cf. Figs.~\ref{Fig:TC-bound}b).

\subsection{In Dynamical Left-Right Symmetric Extension of the 
Top-condensate Model}   

Finally, we consider the 
left-right symmetric extension \cite{tt-lindner} of the top-condensate
model, which postulates the gauge structure 
$~{\cal G}_{\rm LR}=SU(3)_c\otimes SU(2)_L\otimes 
SU(2)_R\otimes U(1)_{B-L}~$
at a high energy scale $\Lambda$.
This model has many attractive features. For example,
parity violation can appear naturally via
the spontaneous symmetry breaking and the known quarks
and leptons fit economically into 
fundamental representations of the gauge group.
A dynamical see-saw mechanism can also be 
realized in this scenario, which
naturally yields the small neutrino masses.

At the compositeness scale $\Lambda$, a set of NJL-type
four-Fermi interactions are generated, which produce a composite Higgs
sector at the lower scale $\mu (<\Lambda )$. 
The symmetry breaking 
pattern occurs via two steps: first, $~{\cal G}_{LR}~$
breaks down to 
$~{\cal G}_{\rm SM}=SU(3)_c\otimes SU(2)_L\otimes U(1)_Y~$ at a scale
$\mu =\Lambda_{\rm R}$; second, the
remaining standard model group $~{\cal G}_{\rm SM}~$ is broken down to 
$~SU(3)_c\otimes U(1)_{\rm em}$~ at the electroweak scale 
of $O$($100$~GeV). 
The composite Higgs sector of this model
contains a scalar bi-doublet $\Phi$, 
two scalar doublets $\chi_L$
and $\chi_R$, and a singlet scalar $\sigma$, with the quantum number 
assignments $(1,2,2,0)$, $(1,2,1,-1)$, $(1,1,2,-1)$ and $(1,1,1,0)$, 
respectively. They are defined as:
{\small
\be
\begin{array}{llll}
\Phi =\dis\left(\begin{array}{cc}
\dis\f{\phi_1^0+v_1}{\sqrt{2}} & \phi_2^+\\
\phi_1^- & \dis\f{\phi_2^0+v_2}{\sqrt{2}} 
                \end{array} \right),~~
& \chi_L=\dis\left(\begin{array}{c}
\dis\f{\chi_L^0+v_L}{\sqrt{2}} \\
\chi_L^-   \end{array} \right),~~
& \chi_R=\dis\left(\begin{array}{c}
\dis\f{\chi_R^0+v_R}{\sqrt{2}} \\
\chi_R^-   \end{array} \right),~~
& \sigma ~,
\end{array} 
\label{eq:LR-Higgs}
\ee
}
\noindent 
where the vacuum expectation value $v_R$ is much larger than the
other VEVs
($v_{1,2}$ and $v_L$) and is responsible for the first step
breaking of the left-right symmetry.  The true VEV ($v\simeq 246$~GeV)
for the EWSB is determined by
\be
v^2=\left(v_1^2+v_2^2\right)+\f{1}{2}\left[\left(v_R^2+v_L^2\right)-
\dis\sqrt{\left(v_R^2-v_L^2\right)^2+\left(4v_1v_2\right)^2}\right]
\simeq v_1^2+v_2^2+v_L^2 ,
\label{eq:LR-VEV}
\ee
where the approximate relation holds, because $~v_R\gg v_{1,2},~v_L$.
Note that  a nonzero $v_L$ (which may be relatively small) implies 
$~v_{12}\equiv \sqrt{v_1^2+v_2^2} < v\simeq 246~$GeV.
Furthermore, the singlet scalar $\sigma$ does not develop VEV.

The relevant Yukawa interactions and mass terms 
for the top-bottom sector 
can be written as \cite{tt-lindner}:
\be
\begin{array}{ll}
{\cal L}_{SF}^{tb} = & 
m_t\tbar t+m_b\bbar b+
\f{1}{\sqrt{2}}\tbar \left(y_1\phi_1^0+y_2\phi_2^0\right) t +
\f{1}{\sqrt{2}}\bbar \left(y_1\phi_2^0+y_2\phi_1^0\right) b \\[0.15cm]
& +\left[ \bbar_L \left(y_1\phi_1^--y_2\phi_2^-\right)t_R+
   \tbar_L \left(y_1\phi_2^+-y_2\phi_1^+\right)b_R+ {\rm h.c.}\right]~,
\\[0.25cm]
& m_t=\left(y_1v_1+y_2v_2\right)/\sqrt{2}~,~~~~
  m_b=\left(y_1v_2+y_2v_1\right)/\sqrt{2}~,
\end{array}
\label{eq:LR-Yukawa}
\ee
which only involve the scalars in the bi-doublet $\Phi$.
To give the correct top mass, $\tanb$ ($\equiv v_1/v_2$)
is constrained to be in the range of $1.3-4.0$.  
The formation of dynamical condensates
or the VEVs of the composite Higgs scalars requires the 
Yukawa couplings $~y_{1,2}~$ to be above their critical value at the
compositeness scale. Consequently, $~y_{1,2}~$ at the weak scale
can be naturally large  [of $\sim O(1)$].
The RG analysis \cite{tt-lindner} indeed shows that for 
$\Lambda$ to be in the range of $10^{5}$ to $10^{19}$~GeV, 
the Yukawa coupling $y_1(\mu )$ varies from about $2.1$ to $1.2$
at the scale $\mu =O(100-1000)$~GeV. 
Since the Yukawa coupling $y_2(\mu )$
satisfies the same RG equation as that of $y_1(\mu )$
(after interchanging $y_1$ and $y_2$) \cite{tt-lindner},
the infrared 
value of $y_2(\mu )$ is also naturally large [of $O(1)$], and
is not sensitive to the boundary condition at the 
compositeness scale.\footnote{
Here, the compositeness scale $\Lambda$
can be as low as $100$~TeV and the left-right breaking scale 
$\Lambda_{\rm R}=v_R$ is around of $O$($10$~TeV) \cite{tt-lindner}.}~
Furthermore, the mass of the neutral $CP$-even and $CP$-odd scalars
Re$(\phi_2^0)$ and Im$(\phi_2^0)$ may be as light as about
$O(100)$~GeV \cite{tt-lindner}.
We thus expect that measuring
the production rate of these light scalar bosons 
via the $\phi b \bar b$ mode at the Tevatron and the LHC can effectively 
test this model.

Before concluding this section, we note that in the three types of
dynamical models discussed above, the relevant composite 
Higgs scalars (having large Yukawa coupling $y_b$) do not couple to 
$\tau^-\tau^+$ mode at the tree level. 
This is in contrast to the case of MSSM
where the lepton-Higgs Yukawa couplings 
(such as $A$-$\tau^+$-$\tau^-$ etc)
are enhanced in the same way as 
that for the down-type quarks in the large
$\tanb$ region. Therefore, further combining the 
$b\bbar \phi (\to \tau^-\tau^+ )$
channel into our analysis would be useful to discriminate the above 
dynamical models from the MSSM\footnote{The $\tau$ Yukawa couplings to
other possible composite scalars are not yet well specified in the
top-condensate/topcolor models, while for the dynamical left-right model
the $\tau$ Yukawa couplings are expected to be naturally small, not much
different from the SM value \cite{tt-lindner,lindner}.}, should a signal
be observed.

\section{CONSTRAINTS ON MSSM PARAMETERS AND IMPLICATIONS FOR 
MODELS OF SOFT-BREAKING OF SUSY}

Supersymmetry (SUSY) is one of the most natural extensions of the SM, 
mainly because of its ability to solve the hierarchy problem,
as well as for its capacity to imitate the current experimental 
success of the SM, despite the plethora of
the introduced new particles and free parameters \cite{susyrev}. 
The Minimal Supersymmetric SM (MSSM) \cite{mssm}
requires a two Higgs doublet extension of the SM
\cite{hhunt} together with the corresponding supersymmetric partners.  
The model includes all renormalizable interactions that respect  
the standard gauge group $SU(3)_C\otimes SU(2)_L\otimes U(1)_Y$ and
supersymmetry. In order to prevent potentially dangerous
baryon and lepton number violating interactions, invariance under
a discrete $R$-parity\footnote{The $R$-parity is
defined in such a way that SM particles are
even under $R$ and their superpartners are odd.} is also required. 
To be compatible with data, supersymmetry has to be broken.
The breaking of SUSY is parametrized by the general set of
soft-breaking terms, which, in principle, should be deduced from a
specific underlying model for SUSY breaking, such as the
supergravity \cite{susyrev} and gauge-mediated \cite{GMSB} models.
In this section, we discuss the potential of
the Tevatron and the LHC to test MSSM via measuring the production 
rate of $\phibb$ mode. We shall also discuss the implication of
this result on various supergravity and gauge-mediated 
models of soft SUSY-breaking.

\subsection{Bottom Yukawa Couplings and the MSSM Higgs Sector} 

In the MSSM, the Higgs couplings to the SM fermions and gauge bosons
involve two new free parameters at the 
tree-level, which are the vacuum angle 
$\beta (\equiv \arctan v_u/v_d)$ and the Higgs mixing angle $\alpha$.
These couplings are shown in Table~\ref{Hcoupling}. 
We see that the MSSM Higgs
couplings to the gauge boson pairs 
are always suppressed relative to that of 
the SM, while their couplings to 
the down(up)-type fermions are enhanced in
the large (small) $\tanb$ region. These enhanced Yukawa couplings are of
great phenomenological importance for the Higgs 
detection and especially for probing
the associated new dynamics in the top-bottom sector.  
Alternatively, we can choose $\tanb$ and the pseudoscalar mass $m_A$ 
as two free parameters. 
Then, at the one-loop level, $\alpha$ can be calculated from
\begin{equation}
\dis\tan 2\alpha 
   =\tan 2\beta\left(m_A^2+m_Z^2\right)
               \left[m_A^2-m_Z^2+\f{\epsilon_t}{\cos 2\beta}\right]^{-1}~,
\label{eq:MSSM-alpha}
\end{equation}
with $~\alpha\in \left(-\pi /2, 0\right)$.~ Here the parameter $\epsilon_t$
represents the dominant top and stop loop corrections which depend on
the fourth power of the top mass $m_t$ and the logarithm of the 
stop mass $M^2_{\tilde t}$:  
\begin{equation}
\dis\epsilon_t= \frac{3 G_F m^4_t}{\sqrt{2}\pi^2 \sin^2 \beta}
\log\left(\f{M^2_{\tilde t}}{m^2_t}+1\right)~.
\label{eq:MSSM-alpha1}
\end{equation}
Note that for large $\tanb$, the bottom and sbottom loop corrections
can also be important. Hence, in our numerical analysis below (cf. Sec.~IVB), 
we have included the complete
radiative corrections with full mixing in the stop and sbottom sectors,
and the renormalization group improvements are also adopted.\footnote{ 
We have used the HDECAY program \cite{hdecay} to compute the Higgs
masses, couplings and decay branching ratios.}
As shown in Table~\ref{Hcoupling}, the $A$-$b$-$\bar b$
coupling has no explicit $\alpha$ dependence.
The bottom Yukawa couplings $y_{bbh}$ and $y_{bbH}$ are 
$\alpha$- and $\beta$-dependent, their magnitudes relative to the SM 
prediction are displayed in Fig.~\ref{fig:YsMs}a 
as a function of $m_A$ for various $\tanb$ values. It shows that for
$m_A$ above $\sim 110$~GeV, the $h$-$b$-$\bar b$
coupling quickly decreases, approaching to the
SM value for all $\tanb$, while the $H$-$b$-$\bar b$
coupling increases for large $\tanb$. 
Therefore, we expect that in the large $\tanb$ region, the production
rate of $Ab\bar b$ or $hb\bar b$ can be large for small $m_A$, while
the rate of $Ab\bar b$ or $Hb\bar b$ are enhanced for large $m_A$.   
Whether the signals of the two Higgs scalars 
($A$ and $h$ or $A$ and $H$) can be experimentally resolved as 
two separate signals 
({\it e.g.,} two bumps in the $b \bar b$ invariant mass distribution)
depends on their mass degeneracy.

The MSSM Higgs boson mass spectrum
 can be determined by taking the second derivative on
Higgs effective potential with respect to the Higgs fields. 
At tree-level, the resulting Higgs masses obey the relations:
$m_h \leq m_Z \cos 2\beta$, $m_Z \leq m_H$, 
$m_h\leq m_A \leq m_H$, and $m_W \leq m_{H^\pm}$.
However, these relations are substantially modified 
by radiative corrections \cite{himrc}. 
Including the important contributions from top and stop
loops, the masses of $h$ and $H$ can be written as: 
\begin{equation}
m^2_{h,H} = \frac{1}{2} \left\{ \left(M^2+\epsilon_t\right) \mp 
\left[ \left(M^2+\epsilon_t\right)^2
-4\epsilon_t\left(m_Z^2\cos^2\beta +m_A^2\sin^2\beta\right)
-4m_Z^2m_A^2\cos^22\beta 
\right]^{1\over 2} \right\},
\end{equation}
where $M^2\equiv m_Z^2+m_A^2$ and the parameter $\epsilon_t$ is defined in 
(\ref{eq:MSSM-alpha1}).
To analyze the Higgs mass degeneracies, we plot the mass differences
$m_A$-$m_h$ and $m_H$-$m_A$ in Fig.~\ref{fig:YsMs}b 
using the complete radiative corrections to the Higgs mass spectrum 
\cite{hdecay}.  We see that for the large $\tanb$ values, 
the pseudoscalar $A$ is about degenerate in mass
with the lighter neutral scalar $h$ below $\sim 120$~GeV and with the 
heavier neutral $H$ above $\sim 120$~GeV. This degeneracy indicates that
the $\phibb$ signal from the MSSM generally contains two
mass-degenerate scalars with similar couplings, and thus
results in a stronger bound on $\tanb$ by about a factor of $\sqrt{2}$.

Finally, we note that in Fig.~\ref{fig:YsMs}, all 
soft-breaking mass parameters were chosen to be 500~GeV. 
Various choices of SUSY soft-breaking parameters 
typically affect these quantities
by about 10\%-30$\%$. To illustrate these effects, we plot in 
Fig.~\ref{fig:YsMs2} the same quantities, but
changing the right-handed stop mass to
$M_{\tilde t}=200$~GeV in (a-b), and in (c-d) we use the
``LEP2~II Scan~A2'' set of SUSY parameters\footnote{
The parameters $m_0$ and $M_2$ are fixed at $1$~TeV, $\mu$ is
chosen to be $-100$~GeV and $m_t=175$~GeV. The scalar trilinear
coupling $A_i$ is fixed at $\sqrt{6}$~TeV, corresponding to
the maximal left- and right-handed top-squark mixing. Detailed 
prescription about this set of parameter scan can be found in 
Ref.~\cite{madison-LEP2,FNAL-LEP2}.}
for comparison. 

Because the MSSM predicts a large bottom quark Yukawa coupling 
for large $\tanb$, and the mass of the lightest neutral scalar has
to be less than $\sim 130$\,GeV, we expect that
the Tevatron and the LHC can test this model via measuring 
the $\phibb$ production rate.
In the following, we shall discuss the range of the
$m_A$-$\tanb$ plane that can be explored at various colliders.
Some models of SUSY soft-breaking predict a large $\tanb$ with
light Higgs scalar(s), and thus predict a large 
$\phibb$ rate. Without observing such a signal, one can put a 
stringent constrain on the model. 
This will also be discussed below.

\subsection{Constraints	on MSSM from $\phibb$
production at Tevatron and LHC}

To use the model-independent result of $K_{min}$ obtained in
Sec.~II to constrain the  $m_A$-$\tanb$ plane in the MSSM, 
one needs to calculate the SUSY Higgs boson masses, 
decay branching ratios, and their couplings to
the bottom quark for a given set of the soft breaking parameters.  
In the following numerical analysis, we use the HDECAY code to include 
the full mixings in the stop/sbottom sector with QCD and electroweak 
radiative corrections \cite{hdecay}.
For simplicity, we assume that the superpartners are all heavy enough 
so that the decays of the Higgs bosons into them are forbidden.
Under this assumption, we find that the 
decay branching ratio of $h \to b
\bar{b}$ is close to one for the relevant region of the parameter space.

As explained above, we combine signals from more than one scalar boson
provided their masses differ by less than $\Delta m_{exp}$, which
is the maximum of the experimental mass resolution (cf. footnote-7)
and the natural decay width of the scalar boson.
Since the results of Sec.~II are given in terms of the
minimal enhancement factor $K_{\min}$ defined in (1),
we need to convert them into exclusion bounds in the $m_A$-$\tanb$
plane of the MSSM, in case that a signal is not found.
The bound on $\tan\beta_{\min}$
(with the possible mass degeneracy included)
can be derived from that on $K_{\min}$ (for a single scalar) by 
requiring
\begin{equation}
\tan^2\hspace*{-0.15cm}\beta~ {\rm BR}(A\to b\bbar) 
+\hspace*{-0.15cm}\sum_{\phi =h,H}\hspace*{-0.2cm}
\dis\theta \left(\Delta m_{\rm exp} -|\Delta M_{A\phi}|\right)
\dis\left(\f{y_b^\phi}{y_b^{\rm SM}}\right)^2{\rm BR}(\phi\to b\bbar)
\geq K_{\min}^2 ~,
\label{eq:MSSM-condition}
\end{equation}
where  $y_b^{\rm SM}$ and $y_b^\phi$ denote the $b$ 
quark Yukawa coupling 
in the SM and the MSSM (with $\phi =h$ or $H$),  respectively.
Inside the argument of the $\theta$-function,  
$\Delta M_{A\phi}$ is the mass difference
between $A$ and $\phi$.
Thus, the equality sign in the above relation 
determines the minimal value
$\tan\beta_{\min}$ for each given $K_{\min}$. 

To estimate the exclusion regions in the $m_A$-$\tanb$ plane, 
a set of soft breaking parameters has to be chosen, which 
should be compatible with the current data from the 
LEP~II and the Tevatron experiments, while not much larger than 1 TeV.
For simplicity, we choose 
all the soft SUSY breaking parameters 
(and the Higgs mixing parameter-$\mu$)
to be 500 GeV as our ``default''
values, i.e., $M_{\rm soft}=500$~GeV.
In Fig.~\ref{fig:Exclusion}a, we show the 
$95\%$~C.L. exclusion contours in the $m_A$-$\tan\beta$ plane 
derived from the measurement of 
$p\bar{p}/pp \to \phibb \to \bbbb$, using this ``default'' set
of SUSY parameters.
The areas above the four boundaries are 
excluded for the Tevatron Run~II 
with the indicated luminosities, 
and for the LHC with an integrated luminosity of 100 fb$^{-1}$.
Needless to say, different choice of SUSY parameters, such as the 
mass and the mixing of the top squarks and the value (and sign) of the
parameter $\mu$, would modify this result. 
To compare the potential
of the Tevatron and the LHC in constraining 
the MSSM parameters via
the $\phibb$ ($\phi =h,A,H$) production 
to that of the LEP~II experiments
via $Z\phi$ and $hA$ production, we consider one of the ``benchmark''
parameter scans discussed in 
\cite{madison-LEP2,FNAL-LEP2}, which is called the
``LEP II Scan~A2''$^{12}$ set. 
For this set of SUSY parameters, the 
LEP~II exclusion contour ~\cite{madison-LEP2,FNAL-LEP2} 
at the 95\% C.L. is displayed in 
Fig.~\ref{fig:Exclusion}b,
for a center-of-mass energy of 
$200$~GeV and an integrated luminosity of
$100$~pb$^{-1}$ per LEP~II experiment.
As shown in Fig.~\ref{fig:Exclusion}b, 
the Tevatron Run~II result, in comparison with the LEP~II result,
can already cover a substantial region
of the parameter space with only a 2 fb$^{-1}$ luminosity.
Thus, detecting the $\phibb$ signal at hadron colliders
can effectively probe the MSSM Higgs sector,
especially for models with large $\tan \beta$ values.
Furthermore, for $m_A \gae 100$ GeV, 
Tevatron Run~II is complementary with LEP~II,
because the latter is not sensitive to that region of parameter space.
The LHC can further probe the MSSM down to $\tanb \sim$ 7~(15) 
for  $m_A < 400~(1000)$ GeV.
This is also shown in Fig.~\ref{fig:ExclusionLHC}
using the ``default'' set of SUSY parameters,
in which 
the region above the upper curve is the discovery contour
at the $5 \sigma$ level
for the LHC with an integrated luminosity of 100 fb$^{-1}$,
and the area above the lower curve can be excluded at $95\%$~C.L,
if a signal is not found.

For completeness, we also present the exclusion contours 
in the $m_h$-$\tanb$ and $m_A$-$\tanb$
planes for both the ``default" and the ``LEP II Scan A2'' sets
of SUSY parameters. 
They are shown in Figs.~\ref{fig:Exclusion2} and 
\ref{fig:Exclusion3}.
Again, we see that the Tevatron Run II and the LHC bounds sensitively
and complementarily cover the MSSM parameter space in contrast with
the LEP II results.
We have also studied the bounds with the ``LEP II Scan A1''  inputs
\cite{madison-LEP2,FNAL-LEP2} and found that the exclusion 
contours from the $\phibb$ production are similar to those with 
``LEP II Scan A2'' inputs.
The most noticeable difference is that the theoretically allowed 
range for $m_h$ becomes smaller by about 10\,GeV in the ``Scan A1'' set,
as compared to the ``Scan A2'' inputs.

Even though our analyses, described above and in Sec.~II, 
are quite different from that of Ref.~\cite{dai}, the final bounds 
at the LHC happen to be in qualitative agreement. 
We also note that our bounds on the $m_A$-$\tanb$ plane
improve considerably the one obtained in Ref. \cite{dressetal},
in which  
the $p{\bar p}\to \phibb \to \tau^+ \tau^- b \bar b$ production
rate at the Tevatron Run~I data
was compared to the MSSM prediction.
Though, we
do not choose to explicitly present projected results for the Tevatron
Run~I data, we encourage our experimental colleagues to pursue this 
analysis
on the existing Run~I data sample, as it seems likely that one could
obtain useful information even with the lower luminosity and collider 
energy of Run~I as well as a somewhat lower $b$-tagging efficiency.

Before concluding this subsection,
we remark upon the effects on our bounds from the
possible radiative corrections to the
$\phibb$ production process.\footnote{This point has also been recently
discussed in Ref.~\cite{stev}.}  As mentioned earlier
in Sec.~II, one of the dominant correction is from the next-to-leading
order (NLO) QCD loops, which are not currently available
for the $\phibb$ signal and background cross sections. 
However, aside from the QCD corrections to the
$\phibb$ vertex (part of that can be simply 
included into the running of the
$\phibb$ Yukawa coupling or the running $b$-mass), there are 
pentagon loops formed by the virtual gluons 
radiated from an initial state quark (gluon) and re-absorbed by the 
final state $b$ quark with the $\phibb$ vertex included in the loop.
Such QCD corrections are not factorizable so 
that a consistent improvement
of our results is impossible before a full NLO QCD analysis is 
completed\footnote{Such a full NLO QCD calculation is beyond the scope
of our current study. A systematic calculation for this is in 
progress \cite{spira}.}. 
Putting aside the complexity of the full NLO QCD
corrections, we briefly comment upon how the
radiative corrections to the running $\phibb$ Yukawa coupling
affect our final bounds.
The well-known QCD correction, Eq.~ (\ref{eq:running}),
alone will reduce the running mass $m_b(\mu )$ by about $40\%$ 
from the scale $\mu = m_b^{\rm pole}\simeq 5$~GeV 
up to the weak scale of $O(200)$~GeV (cf. the
solid curve of Fig.~\ref{fig:mbRun2L}).
This is however not the full story. The complexity comes from the
finite SUSY threshold correction in the large $\tanb$ region
which are potentially large \cite{mbrun2,mbrun,mbrun3}. 
In this case, as shown in Ref.~\cite{mbrun}, the dominant one-loop
SUSY correction contributes to running $b$-mass a finite term so that
$m_b(\mu )$ at the SUSY scale 
$\mu \equiv \mu_R = M_{\rm soft}$ is multiplied by
a constant factor $~1/[1+\Delta_b({\rm SUSY})]~$, which 
appears as a common factor in the bottom Yukawa couplings
of all three neutral Higgs bosons.
For large $\tanb$,
$\Delta_b({\rm SUSY})$ contains the following $\tanb$-enhanced terms
from sbottom-gluino and stop-chargino loops\footnote{We
thank K.~Matchev for discussing his published results in Ref.~\cite{mbrun},  
and to him and W.A.~Bardeen for discussing 
the issue of the running $b$-mass. 
Our convention of the MSSM Higgs parameter $\mu$ 
differs from that of Ref.~\cite{mbrun} by a minus sign.},
\begin{equation}
\begin{array}{l}
\Delta_b({\rm SUSY})~=~ 
\dis\left(\f{\Delta m_b}{m_b}\right)^{\tilde{b}\tilde{g}}
   +\left(\f{\Delta m_b}{m_b}\right)^{\tilde{t}\tilde{\chi}}\\[0.45cm]
=\dis\f{-\mu\tanb}{16\pi^2}\left\{\f{8}{3}g_3^2m_{\tilde{g}}
F\left(m_{\tilde{b}}, m_{\tilde{b}}, m_{\tilde{g}} \right)+
\left[ y_tA_t F\left(m_{\tilde{t}}, m_{\tilde{t}}, \mu \right) -
g_2^2M_2 F\left(m_{\tilde{t}}, m_2, \mu \right)  
\right]\right\},
\end{array}
\label{eq:mbrun}
\end{equation}
with the function $F$ defined as:
$$
F(\sqrt{x},\sqrt{y},\sqrt{z})
=-\dis\f{xy\ln x/y+yz\ln y/z+zx\ln z/x}{(x-y)(y-z)(z-x)} ~.
$$
In these equations, the MSSM Higgs parameter 
$\mu$ should not be confused with 
the usual renormalization scale $\mu_R$.
In (\ref{eq:mbrun}), we have assumed, for simplicity,  mass 
degeneracy for the top and bottom squarks, respectively.
Eq.~(\ref{eq:mbrun}) shows that the SUSY correction to the running $m_b$
is proportional to $\tanb$ and $\mu$. Thus, this correction is 
enhanced for large $\tanb$ and non-negligible in comparison with
the QCD corrections. Also changing the sign of $\mu$ will
vary the sign of the whole correction $\Delta_b({\rm SUSY})$
and implies that the SUSY correction can either increase or decrease
the running $b$-mass at the energy scale around of $O(M_{\rm soft})$.
Normally, when
defining the running coupling using the Collins-Wilczek-Zee (CWZ)
scheme \cite{CWZ}, only the $\mu_R$-dependent contributions are included
while all the $\mu_R$-independent terms are absorbed 
into the corresponding Wilson
coefficient functions. 
However, since the $\mu_R$-independent contribution
$\Delta_b({\rm SUSY})$ is not small for 
large $\tanb$ and a full NLO SUSY
calculation is not yet available, 
we include $\Delta_b({\rm SUSY})$ to 
define an ``effective'' running coupling/mass of $b$-quark
even below the SUSY threshold scale $M_{\rm soft}$.
This could give a rough estimate on the large
SUSY loop corrections from the $\phi$-$b$-$\bar b$ vertex. 
Obviously, when $\mu_R$ is much smaller than
$M_{\rm soft}$, the CWZ scheme should be used. 
Hence,
in Fig.~\ref{fig:mbRun2L}, we only show the ``effective'' running mass
$m_b(\mu )$ down to about $100$~GeV which is the relevant energy scale
and the mass scale ($\sim M_{\rm Higgs}$) considered in this paper.
Fig.~\ref{fig:mbRun2L}  
illustrates that due to the SUSY correction the 
``effective'' running $b$ mass at
the weak scale can be either larger or smaller than the 
SM QCD running value ($\sim 3$~GeV), depending on the choice of
the sign of $\mu$ parameter (and also other soft-breaking parameters
such as the trilinear coupling $A_t$ and masses of the gluino, gaugino,
stop and sbottom).  
For the ``default'' set of SUSY parameters used in our analysis,
all the soft-breaking parameters are set to $500$~GeV for simplicity.
It happens to be the case that
the SM QCD and SUSY corrections nearly cancel each other so that
the ``effective'' running mass $m_b$ is very close to the 
pole mass value ($\sim 5$~GeV) for the scale above $\sim 100$~GeV 
(cf. upper curve of Fig.~\ref{fig:mbRun2L}).
For comparison, in our analysis using 
the ``LEP~II Scan~A2'' set of SUSY parameters, 
the SM QCD and SUSY corrections do not cancel and tend
to reduce the ``effective'' running $b$-mass or the 
Yukawa coupling $y_b(\mu )$
which results in a weaker bound for the Tevatron
Run~II and the LHC, 
as shown in Fig.~\ref{fig:Exclusion}b.\footnote{
A detailed analysis of these effects at the Tevatron Run~II is
currently underway \cite{stev}.} 
The difference in the exclusion contours shown in 
Figs.~\ref{fig:Exclusion}a and~\ref{fig:Exclusion}b,
derived from the measurement
of the $\phibb$ production rate at hadron colliders,
is mainly due to the difference in the ``effective'' 
running coupling or mass (including the QCD and
SUSY corrections), as described above.\footnote{
From Fig.~\ref{fig:Exclusion}, it is also clear that, 
for the LHC bounds, the SUSY correction has much less 
impact since the relevant $\tanb$ values become much lower,
around of $O(2-15)$.}~
We therefore conclude that a full NLO QCD calculation 
is important for a consistent improvement
of our current analysis.

\subsection{Interpretation of results for
Models of soft breaking parameters}

The MSSM allows for a very general set of soft SUSY-breaking
terms and thus is specified by a large 
number of free parameters ($\simeq 124$ \cite{mssm}), 
though only a complicated subset
of this parameter space is consistent with all current
experimental results.  It is therefore important to understand
the mechanism of supersymmetry breaking (which presumably
occurs at a high energy scale \cite{kanet}) and to predict
the soft parameters at the weak scale from an underlying model.
Many alternative ideas about how supersymmetry might be broken, and how
this will result in the low energy soft breaking parameters exist in the
literature, including the supergravity inspired (SUGRA) models and
gauge-mediated SUSY breaking (GMSB) models.  In this section we
examine the sensitivity of the $\phibb$ process to probe 
a few models of SUSY breaking, concentrating for the most part on
the popular SUGRA and GMSB models.  However, there are also 
other interesting ideas to which the $\phibb$ process may provide
interesting information,
because these models naturally prefer a large $\tan \beta$.  
A few examples include the SO(10) 
grand unification theories \cite{sotengut}
(SO(10) GUTs); the infrared fixed-point scenario
\cite{irfixedp}; and also a scenario with compositeness,
the ``more minimal supersymmetric SM" \cite{moremssm}.

\subsubsection{Supergravity Models with large $\tan\beta$}

The supergravity inspired (SUGRA) models \cite{sugrarev}
incorporate gravity in a natural manner, and solve the problem 
of SUSY breaking through the introduction of a hidden sector, 
which breaks SUSY at a very high scale [$\sim O(10^{11})$ GeV] 
and interacts with the MSSM fields only gravitationally.
This model offers an exciting glimpse into the possible connection
between the heavy top quark and the EWSB by allowing radiative breaking
of the electroweak symmetry, in which the large top quark Yukawa
coupling can drive one of the Higgs masses negative at energies
$\sim m_Z$.  In the limit of large $\tan \beta$ the bottom and tau
Yukawa couplings can also play an important role \cite{dressnoji}.

Under the assumption that the gravitational
interactions are flavor-blind, this model
determines the entire SUSY spectrum
in terms of five free parameters (at the high energy scale of the
SUSY breaking) including a common scalar mass
($\tilde m_0$), a common gaugino mass ($M$), a common scalar
tri-linear interaction term ($A$), $\tan \beta$, and the sign of the
Higgs mixing parameter (${\rm sgn}(\mu)$).  The weak scale particle
spectrum can then be determined by using the renormalization group
analysis to run the sparticle masses from the high scale to the weak
scale.

Though large $\tan \beta$ is not required by the minimal SUGRA model,
it can naturally be accommodated, as demonstrated in 
\cite{dressnoji,sugrtbg},   where it was found that
a large $\tan \beta$ also generally requires that the pseudoscalar
Higgs mass be light ($m_A \leq$ 200 GeV for $\tan \beta \geq 30$),
because the enhanced $b$ and $\tau$ Yukawa couplings act through the
renormalization group equations to reduce the down-type Higgs mass
term at the weak scale, thus resulting in a light Higgs spectrum.
Since the importance of the $b$ and $\tau$ effects in the RG
analysis increases
with larger $\tan \beta$, as $\tan \beta$ increases the resulting
$m_A$ decreases, making the large $\tan \beta$ scenario in the
SUGRA model particularly easy to probe through
the $\phibb$ process.
From the limits on the $m_A$-$\tan \beta$ plane derived above in
Section IVB, it thus seems likely that from the data of the
Tevatron Run II with 2 \ifb of integrated luminosity, a large
portion of the minimal SUGRA model with $\tan \beta \geq 20$
may be excluded.

\subsubsection{Gauge-mediated SUSY Breaking Models with 
Large $\tan\beta$}

Models with GMSB break SUSY at a scale which is typically much lower
than that present in the SUGRA models.  
The supersymmetry is generally
broken in a hidden sector which directly couples to a set of messenger
chiral superfields.  This induces a difference in mass between 
the fermion and
scalar components of the messenger fields, which in turn generates
masses for the gauginos and sfermions of the MSSM fields via loops
involving the ordinary gauge interactions \cite{dinetal,gmmrev}.
A generic feature of this scenario is that because of the relatively
low scale of SUSY breaking, the gravitino acquires a much smaller mass
than in the SUGRA scenarios, and is generally the lightest
supersymmetric partner (LSP).
While specific models of GMSB vary as to their assumptions and relevant
parameters, generally what must be assumed is the field content of the
messenger sector (including transformation properties under the gauge
group and number of multiplets in the theory) and the scale at which
SUSY is broken in the hidden sector.

The minimal GMSB models can also result in a radiative breaking of the
EWSB, through the renormalization group evolution of the Higgs masses
(driven by the large top Yukawa coupling)
from the effective SUSY breaking scale to the weak scale.  As in 
the SUGRA model case, this
evolution can drive the mass term of the up-type Higgs
negative at the weak scale, thus breaking the electroweak symmetry.  
In fact, because the effective SUSY breaking
scale in a GMSB model is typically much lower than in the SUGRA model
(and thus closer to the weak scale),
in order for the proper EWSB to occur, it was demonstrated in 
\cite{babuetal} that a large $\tan \beta$ is required (about $30$-$40$).
However, because the effective SUSY breaking scale is typically
much lower than in the SUGRA model, the large effects of the $b$ and
$\tau$ on the Higgs mass running do not reduce the Higgs spectrum to the
degree that occurs in the SUGRA model with large $\tan \beta$,
and thus result in a heavier $A$ with mass
of about 400 GeV.  So, this
particular model would only be explored through the $\phibb$ process
at the LHC.  However, more general analyses of the GMSB scenario 
\cite{baggetal,baertal,ratsarid,borzumati},
introducing more degrees of freedom in the messenger
sector than the minimal model, can allow for 
large $\tan \beta$ and relax
$m_A$ to be as low as about 200 GeV.  
Thus, through the $\phibb$ production these more general 
models can be first 
probed at the Tevatron for the relevant mass range, 
and then largely explored at the LHC.

\section{CONCLUSIONS}

It remains a challenging task to determine the underlying 
dynamics of the 
electroweak symmetry breaking and the flavor symmetry breaking.
Either fundamental or
composite Higgs boson(s) may play a central role in the 
mass generation of the weak gauge bosons and the fermions.  
The heavy top quark, with a mass on
the same order as the scale of the electroweak symmetry breaking, 
suggests that the top quark may play 
a special role in the mechanism of mass generation.  
In this work, we have shown that in the typical 
models of this type, the bottom quark,
as the weak isospin partner of the 
top quark, can also participate in the dynamics of 
mass generation, and serves as an effective probe of the possible new 
physics associated with the Higgs and top sectors.
 
We have presented a model-independent analysis on Higgs boson
production in
association with bottom quarks, via the reaction
$\ppbar/pp \to \phibb \to \bbbb$, at the Tevatron Run~II
and the LHC. We have computed the QCD $\bbbb$ and $\bbjj$ backgrounds,
and the electroweak $\Zbb$ background to illustrate that it is possible
to extract the signal from these large backgrounds by employing a
suitable search strategy.   
The scale and the PDF dependencies in our signal and
background calculations are also examined and  
they indicate that the NLO QCD corrections to the
production rate could be large, and thus including the complete NLO
QCD corrections in future improvements will be very useful.
Using the complete tree level calculation with an estimated QCD
$k$-factor of 2, we derive the exclusion contour for the enhancement
factor (in the coupling of $\phibb$ relative to that of the SM)
versus the Higgs mass $m_\phi$ at the $95\%$~C.L., assuming 
a signal is not found.

We apply these results to analyze the constraints on the parameter
space of both the composite models and 
the MSSM (in the large $\tanb$ region) 
with naturally large bottom Yukawa couplings. 
For the composite Higgs scenario, 
we first consider the two-Higgs-doublet extension of 
top-condensate model and then analyze the topcolor
model, where the $b$ quark Yukawa couplings are naturally large due
to the infrared quasi-fixed-point structure and the particular boundary
conditions for $(y_b,y_t)$ at the compositeness scale. 
Our analysis shows
that the Tevatron Run~II with a 2~fb$^{-1}$ of
luminosity can exclude the
entire parameter space of the simplest two-Higgs-doublet 
extension of top-condensate model,
if a signal is not found.
For the topcolor model, the Tevatron Run~II 
is able to detect the composite Higgs
$h_b$ or $A_b$ up to $\sim 400$~GeV and the LHC can extend the
mass range up to $\sim 1$~TeV.
Similarly, this production mechanism can be 
used to effectively test the dynamical 
left-right symmetric model.  

To confirm the MSSM, it is necessary to 
detect all the predicted neutral 
Higgs bosons $h,~H,~A$ and the charged scalars $H^\pm$.
From LEP~II, depending on the choice 
of the MSSM soft-breaking parameters,
the current $95\%$~C.L. bounds on the masses of the MSSM
Higgs bosons are about 70~GeV for both the $CP$-even scalar $h$ and
the $CP$-odd scalar $A$ \cite{himlim}. It can be improved at LEP~II
with higher luminosity and maximal energy, but the bounds on the
Higgs masses will not be much larger than $\sim m_Z$ for an   
arbitrary $\tanb$ value. The $Wh$ and $WH$ associated production at
the Tevatron Run~II can further improve these bounds, if a signal
is not observed. At the LHC, a large portion of parameter space can be 
tested via $pp\to t\bar t +h(\to \gamma \gamma )+X$, and
$pp\to h(\to ZZ^\ast )+X$, etc \cite{kunstz,gunorr}. 
A future high energy $e^+e^-$ collider will fully test
the MSSM Higgs sector \cite{eehiggs,eehiggs2} through the
reactions $ e^+ e^- \to Z+h(H),\, A+h(H), \, H^+ H^-$, etc. 
In this paper, we demonstrate that studying the $\phibb$
channel at hadron colliders can further improve our knowledge on MSSM.
The exclusion contours on the $m_A$-$\tanb$ plane of the MSSM
shows that the Tevatron and the LHC 
are sensitive to a large portion of the parameter space
via this mode.
It therefore provides a complementary   
probe of the MSSM Higgs sector in comparison with that from LEP~II.
The implications of these bounds in the parameter space
on both the supergravity and the gauge-mediated SUSY
breaking models are further discussed.
We find that the $\phibb$ process can effectively test 
the  models with either scheme of the SUSY soft-breaking
in the large $\tanb$ scenario.

In conclusion, the stringent constraints obtained by
studying the associated $\phibb$ production at hadron colliders
for both the composite and the supersymmetric
models show the utility of this search mode.
However, much work remains to be done.  For example, 
a $b$-trigger would be essential for this analysis.
The CDF group at the Fermilab has demonstrated that
it is possible to detect events with four or more jets and 
two or more $b$-tags \cite{cdfwh},
which can be significantly improved with 
the implementation of $b$-trigger at the Run~II of the Tevatron
\cite{run2btag}.
However, the large QCD 4-jet background rate at the LHC can
potentially make triggering on the 
$b {\bar b} \phi(\rightarrow b {\bar b})$ 
events difficult, though 
it is expected that a $b$-trigger would become more efficient for a 
heavier (pseudo-)scalar $\phi$. (This is because the $b$-jets from 
the decay of a heavier $\phi$ are more energetic, and 
its QCD background rate drops rapidly as a function of the 
transverse momentum of the triggered jet.)
We hope that the interesting results
afforded by studying this channel will stimulate interest 
of our experimental colleagues in working on these problems.

We have found that the $\phibb$ process 
complements Higgs searches in other channels, and 
thus it is expected
that experimental searches for this signature at the Tevatron Run~II
(and possibly beyond) and the CERN LHC will provide interesting
and important information about the mechanism of 
the electroweak symmetry
breaking and the fermion mass generation.

\vspace{0.5cm}
\noindent
{\bf Acknowledgments}\\[0.25cm]
We thank W.A.~Bardeen, E.L.~Berger, B.A.~Dobrescu, K.T.~Matchev, 
C.~Schmidt, M.~Spira, W.K.~Tung, R.~Vilar and H.~Weerts for useful 
discussions, and C. Rembser for providing us the updated LEP~II data
on the MSSM.
We are also indebted to C.~Hill for discussing the topcolor model
and M.~Lindner for discussing the dynamical left-right symmetric model.
We are grateful to S.~Mrenna for help on checking
Monte Carlo simulations and to him and M.~Carena for
conversations on the MSSM phenomenology.
CB and HJH thank the Theory Division of FermiLab for the invitation and
hospitality. CB, HJH and CPY are supported in part by the U.S.~NSF grant
PHY-9507683, and
JLDC and CPY by the CONACYT-NSF international agreement. 
TT's work was done at Argonne National Lab, HEP division, 
and was supported in part by the U.S. DOE Contract W-31-109-Eng-38.


\newpage
\begin{table}[htbp] 
\begin{center}
\begin{tabular}{c||cccc}
Process & Acceptance Cuts   & $p_T$ Cuts & $\Delta R$ Cut & $\Delta M$ 
Cut\\ \hline
$\phibb$  & 4923            & 1936       & 1389       & 1389       \\
$\Zbb$  & 1432              & 580        & 357        & 357        \\
$\bbbb$ & $5.1 \times 10^4$ & 3760       & 1368       & 1284       \\
$\bbjj$ & $1.2 \times 10^7$ & $1.5 \times 10^6$ & 
$6.3 \times 10^5$ & $5.9 \times 10^5$ \\
\end{tabular}
\end{center} 
\vskip 0.08in 
\caption{The signal and background events for 2~\ifb
of Tevatron data, assuming
$m_\phi = 100$\,GeV, $2 \Delta m_\phi = 26$\,GeV, and $K = 40$
after imposing the acceptance cuts, $p_T$ cuts,
and reconstructed $m_\phi$ cuts described in the text.
(A $k$-factor of 2 is included in both the signal and the background 
rates.)}
\label{cutstab}
\end{table}

\begin{table}[htbp] 
\begin{center}
\begin{tabular}{c||cc}
$m_\phi$ (GeV) & $p^{(1)}_T$ Cut & $p^{(2)}_T$ Cut \\ \hline
75             & 35              & 25              \\
100            & 50              & 30              \\
125            & 60              & 35              \\
150            & 70              & 45              \\
175            & 85              & 55              \\
200            & 90              & 60              \\
250            & 125             & 80              \\
300            & 150             & 105             \\
350            & 175             & 190             \\
400            & 190             & 120             \\
500            & 245             & 160             \\
800            & 390             & 260             \\
1000           & 500             & 320             \\
\end{tabular}
\end{center} 
\vskip 0.08in 
\caption{The optimal $p^{(1)}_T$ and $p^{(2)}_T$
cuts for isolating a Higgs boson
of mass $m_\phi$ from the QCD $\bbbb$ background.}
\label{ptcutstab}
\end{table}

\begin{table}[htbp] 
\begin{center}
\begin{tabular}{c||ccc}
~Process     & 2 or more $b$-tags & 3 
or more $b$-tags & 4 $b$-tags  \\ \hline
$\phibb$     & 1139              & 660           & 180              \\
$\Zbb$       & 293               & 170           & 46               \\
$\bbbb$      & 1054              & 610           & 166              \\
$\bbjj$      & $1.2 \times 10^5$ & 2141          & 4   
             \\ \hline
Significance & 3.3               & 12.21         & 12.25            \\
\end{tabular}
\end{center} 
\vskip 0.08in 
\caption{The signal and background events for 2~\ifb
of Tevatron data, assuming
$m_\phi = 100$\,GeV, $2 \Delta m_\phi = 26$\,GeV, and $K = 40$ 
for two or more,
three or more, or four $b$-tags, 
and the resulting significance of the signal. }
\label{btagtab}
\end{table}

\begin{table}[htbp] 
\begin{center}
\begin{tabular}{c||cccc}
~$m_\phi$~(GeV)& \multicolumn{2}{c}{Tevatron}  & \multicolumn{2}{c}{LHC} \\
               & $N_S$        & $N_B$     & $N_S$       & $N_B$     \\ \hline
 75            & 583          & 640       & 3.4 $\times 10^6$ &
 4.8 $\times 10^6$  \\
 100           & 180          & 216       & 2.0 $\times 10^6$ &
 3.0 $\times 10^6$  \\
 150           & 58           & 92        & 9.2 $\times 10^5$ &
 1.2 $\times 10^6$  \\
 200           & 17           & 31        & 4.2 $\times 10^5$  &
 5.6 $\times 10^5$  \\
250            & 4.8          & 8.8       & 1.9 $\times 10^5$  &
 2.0 $\times 10^5$  \\
 300           & 1.3          & 2.1       & 83000       & 70000     \\
 500           &              &           & 12000       & 5700      \\
 800           &              &           & 1500        & 406       \\
 1000          &              &           & 407         & 70        \\
\end{tabular}
\end{center} 
\vskip 0.08in 
\caption{Event numbers of signal ($N_S$), for one Higgs boson, and 
background ($N_B$) 
for a 2~\ifb of Tevatron data and a 100~\ifb of LHC data, for various
values of $m_\phi$, after applying the cuts described in the
text, and requiring 4 $b$-tags. An enhancement of 
$K = 40$ is assumed for the signal, though the numbers may be simply
scaled for any $K_{\rm new}$ by multiplying by $(K_{\rm new}/40)^2$.}
\label{eventnumtab}
\end{table}

\begin{table}[Hcoupling] 
\begin{center}
\begin{tabular}{c||ccc}
\\[-0.3cm]
Higgs &  
$A\hspace*{1.7cm}$  &  
$h\hspace*{1.7cm}$  &  
$H\hspace*{1.7cm}$  \\ [0.25cm]
\hline\hline \\ 
$y_{U}/y_{U}^{\rm SM}$ & 
$\cotb\hspace*{1.5cm}$ & 
$\cosa /\sinb \hspace*{1.5cm}$ & 
$\sina /\sinb\hspace*{1.5cm}$ \\ [0.25cm]
\hline \\ 
$y_{D}/y_{D}^{\rm SM}$ & 
$\tanb\hspace*{1.5cm}$ &  
$-\sina /\cosb \hspace*{1.5cm}$ & 
$\cosa /\cosb\hspace*{1.5cm}$\\[0.25cm]
\hline\\ 
$~~g_{\phi VV}/y_{\phi VV}^{\rm SM}~~$ & 
$0\hspace*{1.5cm}$ &  
$\sinba\hspace*{1.5cm}$ & 
$\cosba\hspace*{1.5cm}$\\ [0.2cm]
\end{tabular}
\end{center} 
\vskip 0.08in 
\caption{
Comparison of the neutral MSSM Higgs couplings to up-type ($U=u,c,t$) 
and down-type ($D=d,s,b; e,\mu , \tau$) fermions and 
to the gauge-boson ($V=W,Z$) pairs. The ratios to the 
corresponding SM couplings are shown, which
are determined by angles $\beta$ and $\alpha$ at the tree-level.
}
\label{Hcoupling}
\end{table}


\newpage
\begin{figure}
\centerline{\hbox{
\psfig{figure=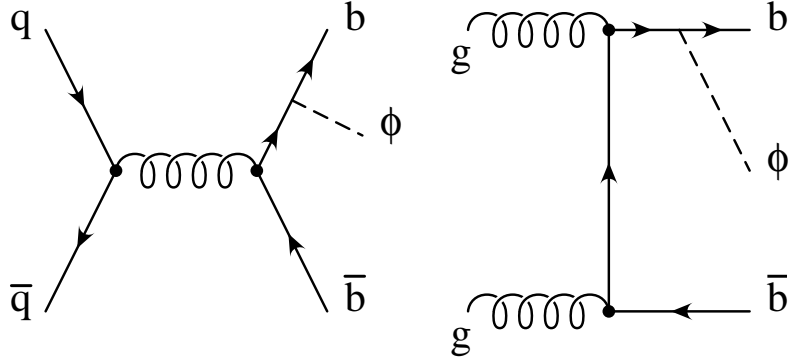,height=2.0in}}}
\vskip 0.1 in
\caption{Representative leading order Feynman diagrams for 
$\phibb$ production
at a hadron collider.  The decay $\phi \to b \bar{b}$ is not shown.}
\label{sigfeynfig}
\end{figure}
\begin{figure}
\centerline{\hbox{
\psfig{figure=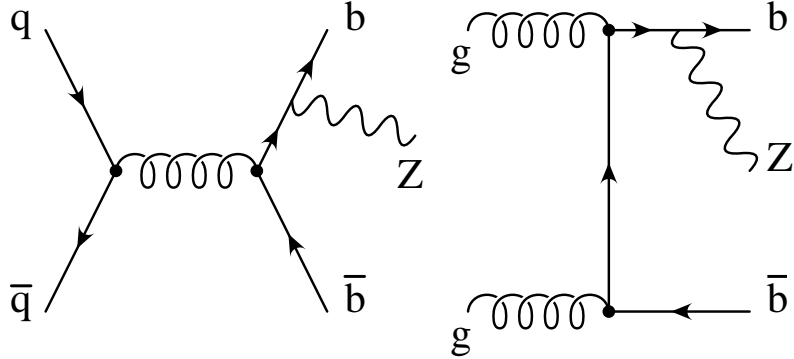,height=2.0in}}}
\vskip 0.1 in
\caption{Representative Feynman diagrams for leading 
order $\Zbb$ production
at a hadron collider.  The decay $Z \to b \bar{b}$ is not shown.}
\label{zbbfeynfig}
\end{figure}

\newpage
\begin{figure}
\centerline{\hbox{
\psfig{figure=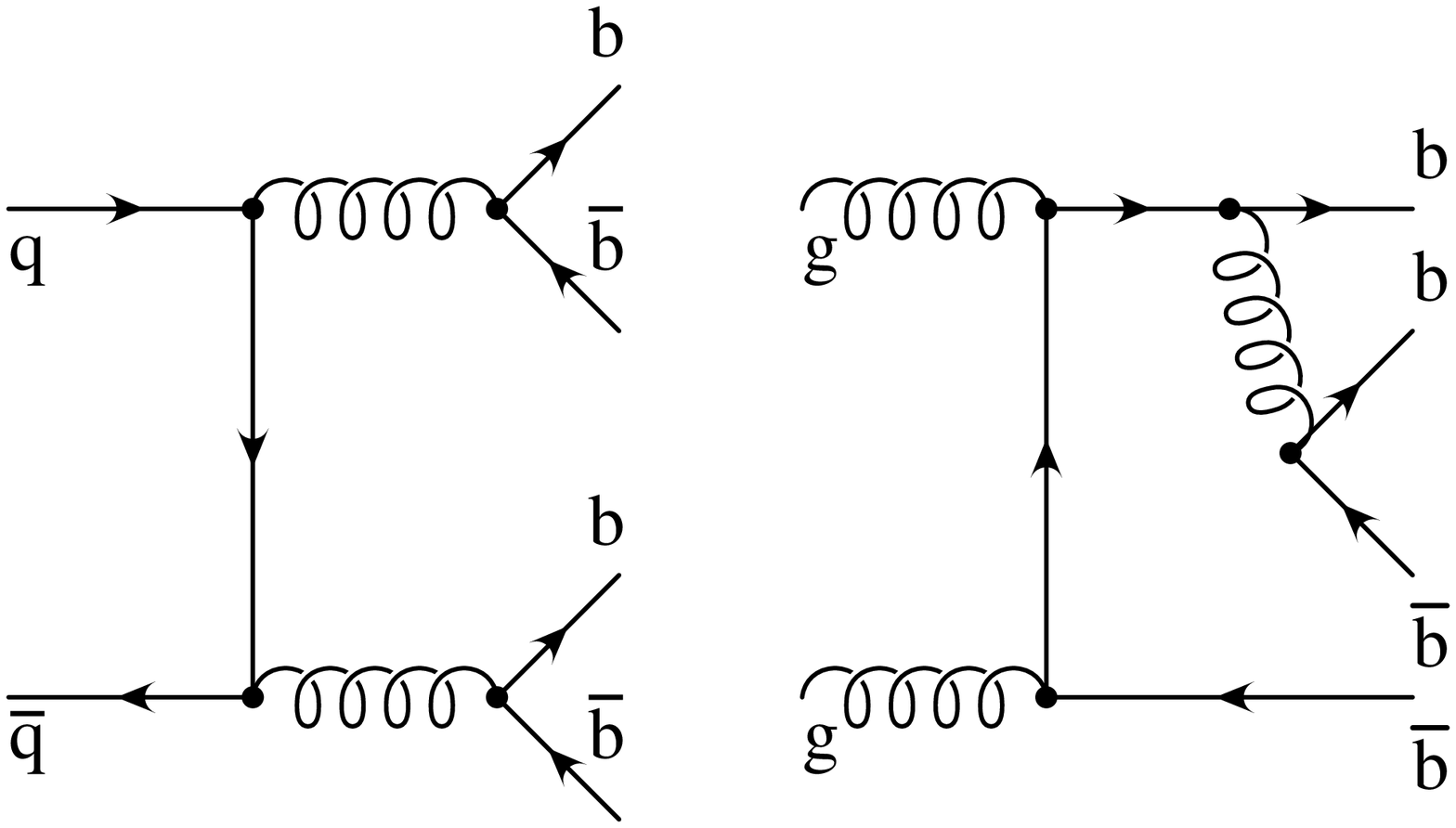,height=2.3in}}}
\vskip 0.1 in
\caption{Representative leading order Feynman diagrams for QCD
$\bbbb$ production at a hadron collider.}
\label{bbbbfeynfig}
\end{figure}
\begin{figure}
\centerline{\hbox{
\psfig{figure=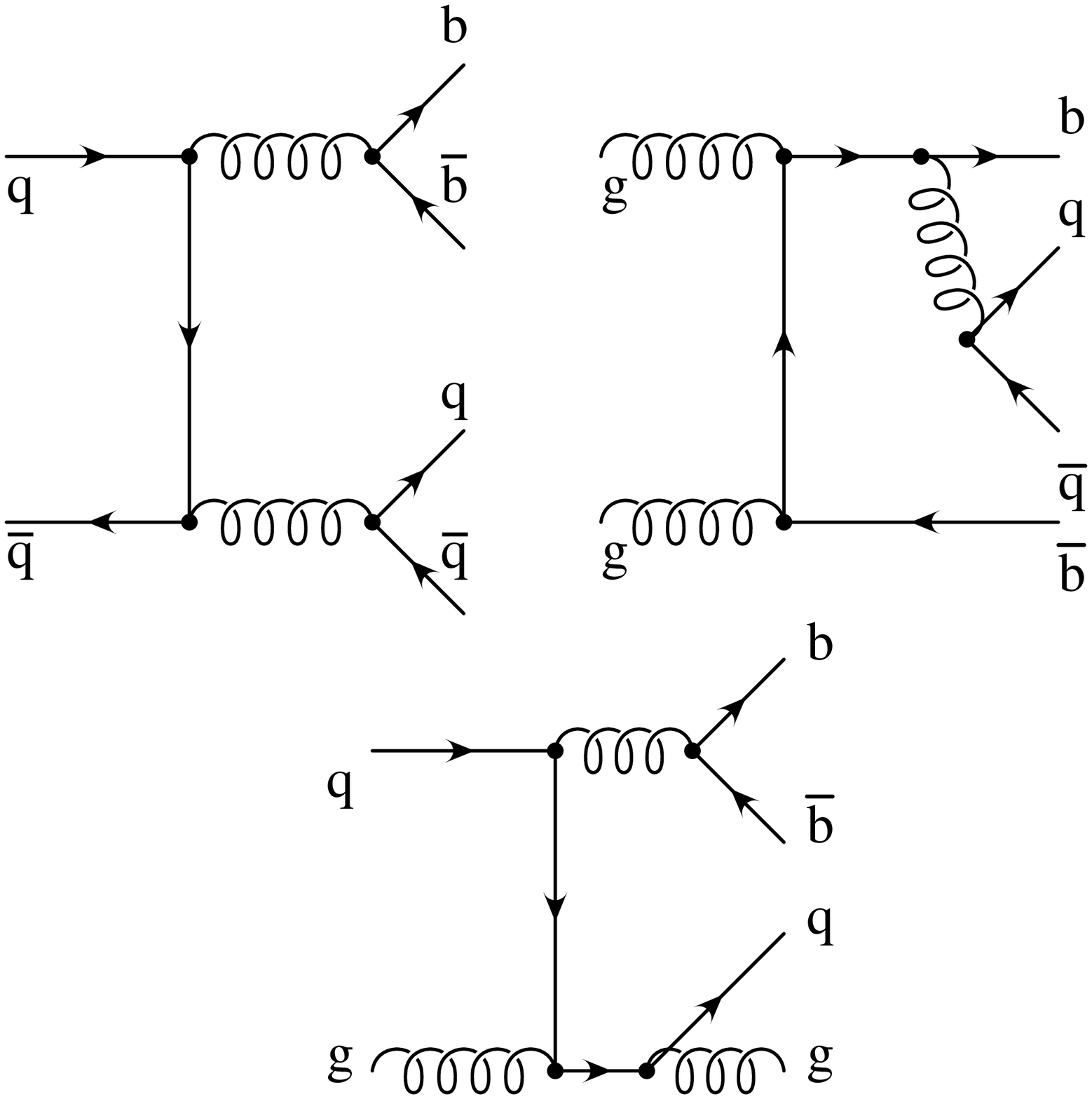,height=4.0in}}}
\vskip 0.1 in
\caption{Representative leading order Feynman diagrams for 
QCD $\bbjj$ production at a hadron collider.}
\label{bbjjfeynfig}
\end{figure}

\newpage
\begin{figure*}[t]
\begin{center}
\vspace*{-0.8cm}
\begin{tabular}{cc} 
\psfig{figure=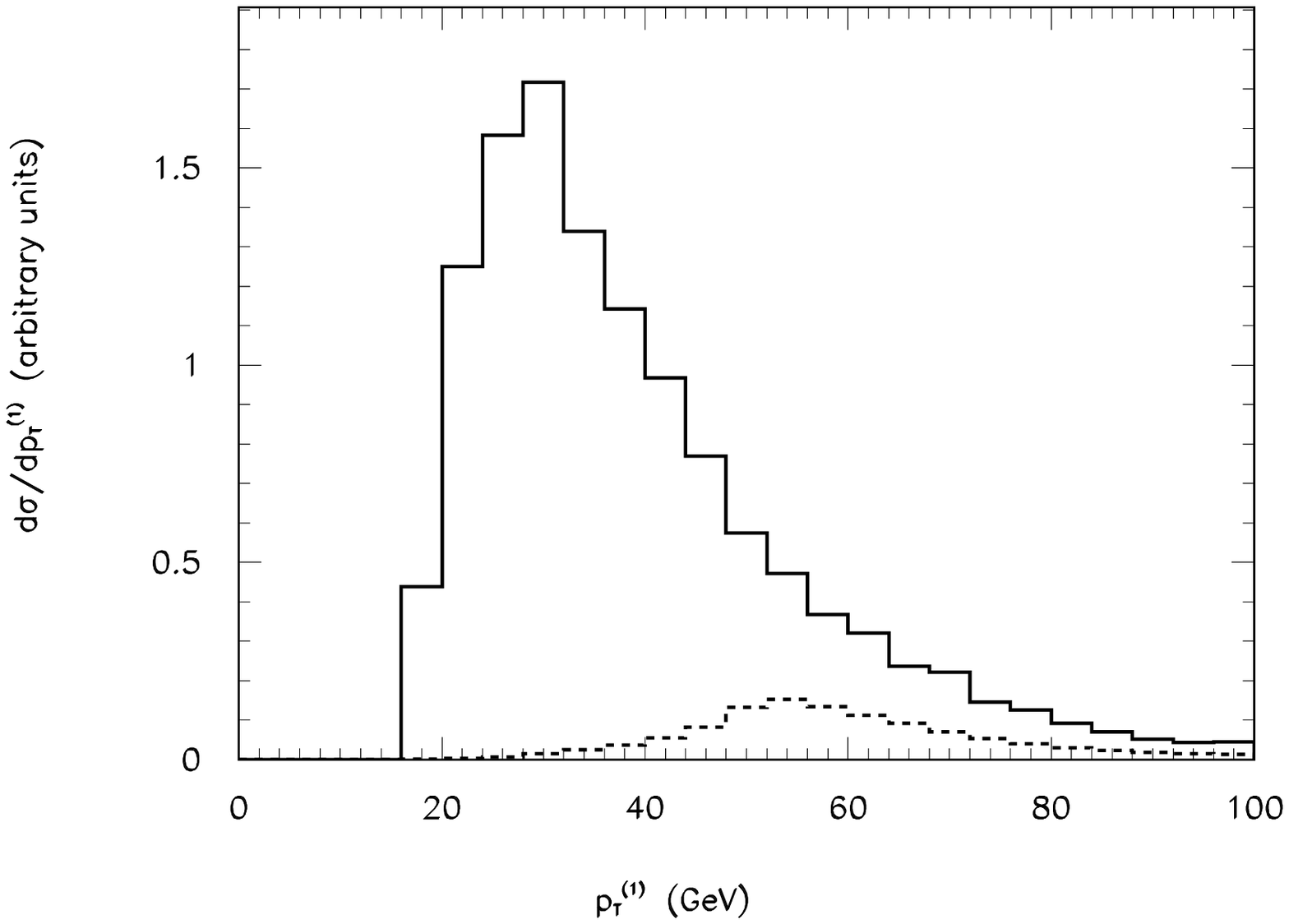,width=16cm,height=3.75in} 
\\[-3.3in]\hskip 5.9in (a)\\[3.0in] \\[-1.4cm]
\psfig{figure=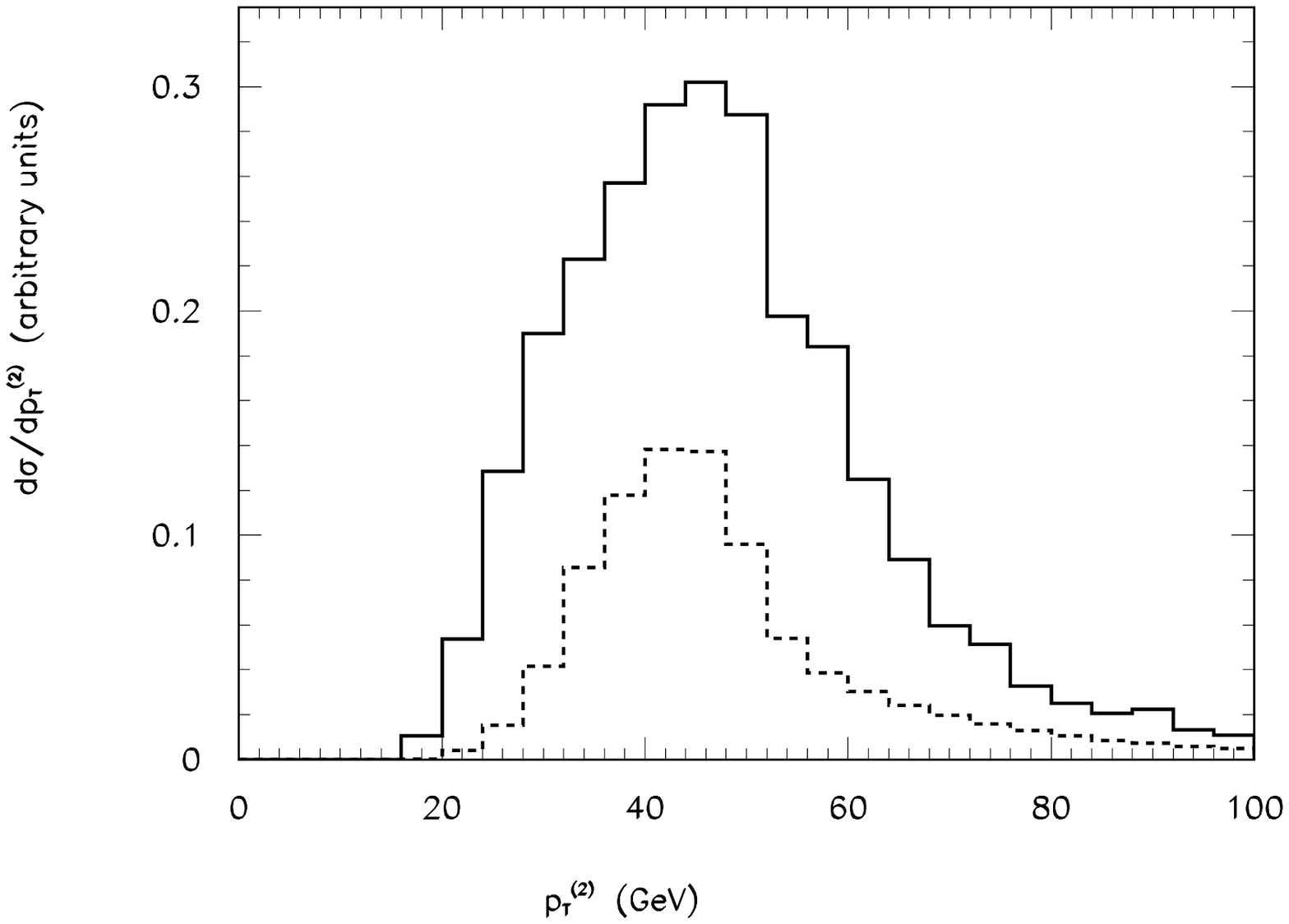,width=16cm,height=3.75in} 
\\[-3.3in]\hskip 5.9in (b)\\[2.8in]
\end{tabular}
\end{center}
\caption{
Figure (a) shows
the distribution of the QCD $\bbbb$ background (solid
curve) and $K = 40$ $\phibb$ signal (dashed curve)
cross sections in $p^{(1)}_T$ at the Tevatron Run II
after the acceptance cuts.  
[$K=y_b/(y_b)_{\rm SM}$, cf. Eq.~(1).]
Figure (b) presents
the distribution of $p^{(2)}_T$ at the Tevatron Run II
after applying the cut to $p^{(1)}_T$,
illustrating the utility of asymmetric cuts on $p^{(1)}_T$ and 
$p^{(2)}_T$ in extracting the $\phibb$ signal from the QCD background.}
\label{fig:pts}
\end{figure*}

\newpage
\begin{figure}
\begin{center}
\vspace*{-1.5cm}
\begin{tabular}{cc} 
\psfig{figure=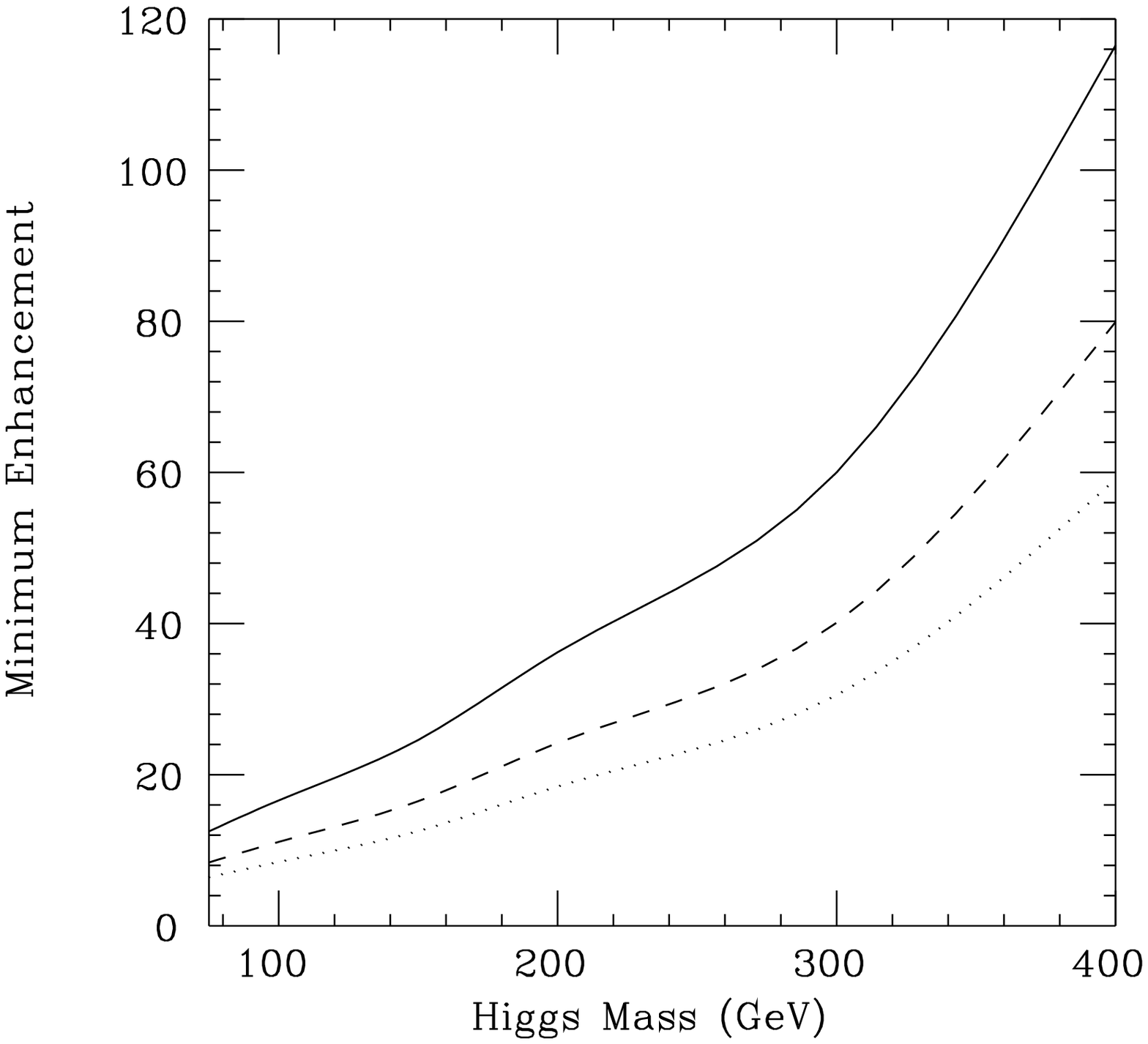,width=16cm,height=3.75in} & 
\\[-3.5in]\hskip 5.9in (a)\\[3.3in] \\[-1.4cm]
\psfig{figure=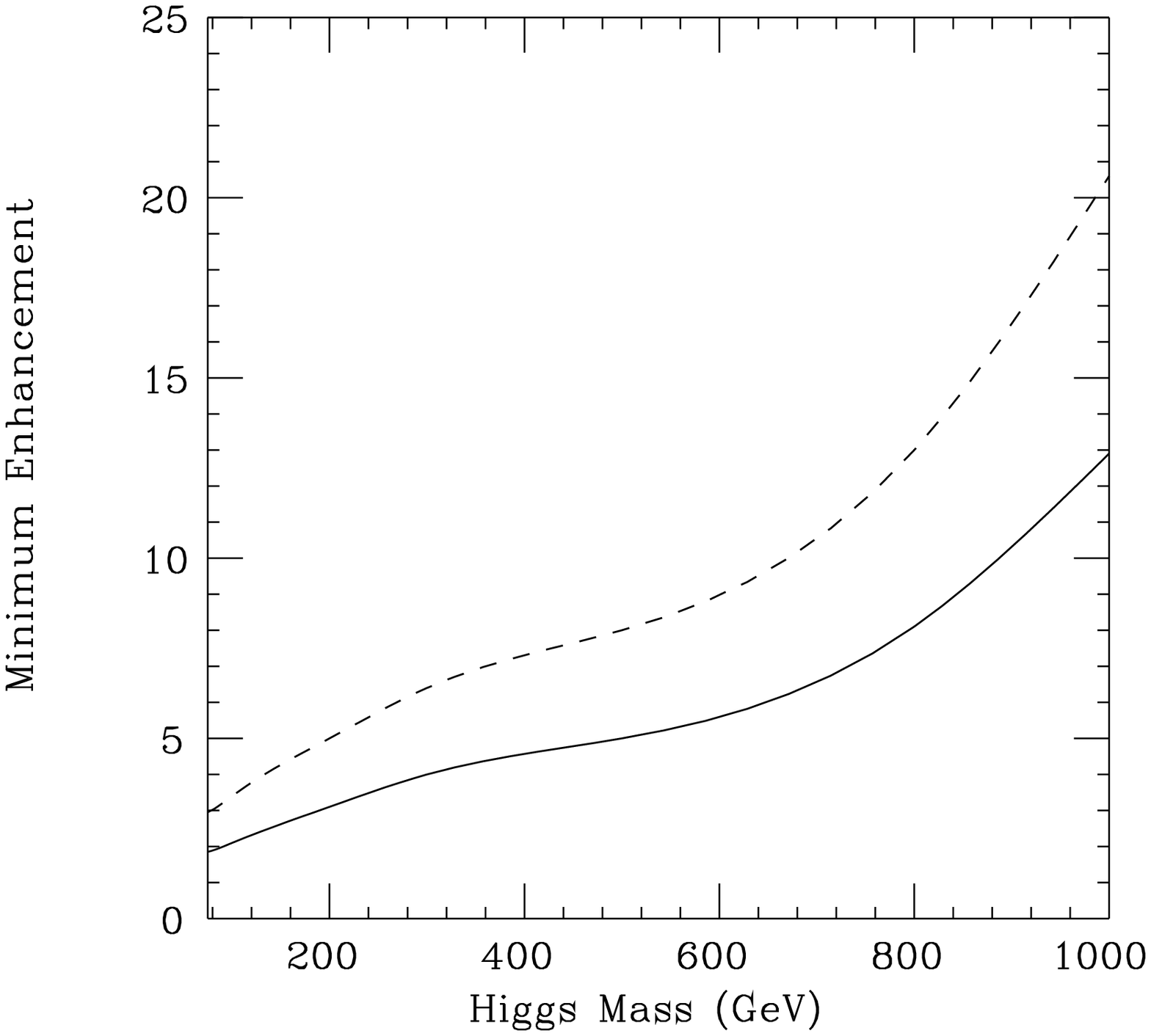,width=16cm,height=3.75in} & 
\\[-3.5in]\hskip 5.9in (b)\\[3.3in]
\end{tabular}
\end{center}
\vspace*{-0.5cm}
\caption{(a). The model-independent minimum enhancement factor, $K_{\min}$, 
excluded at $95\%$ C.L. as
a function of scalar mass ($m_\phi$) for the Tevatron Run II with 2 \ifb
(solid curve), 10 \ifb (dashed curve) 
and 30 \ifb (dotted curve).  
         (b). The same factor, $K_{\min}$, 
excluded at $95\%$ C.L. (solid curve) and discovered at
$5\sigma$ (dashed curve)
as a function of $m_\phi$ for the LHC with 100 \ifb.
In the above, the natural width of the scalar ($\Gamma_\phi$)
is assumed to be much smaller than the experimental mass resolution.}
\label{kminfig}
\end{figure}

\newpage
\begin{figure}
\centerline{\hbox{
\psfig{figure=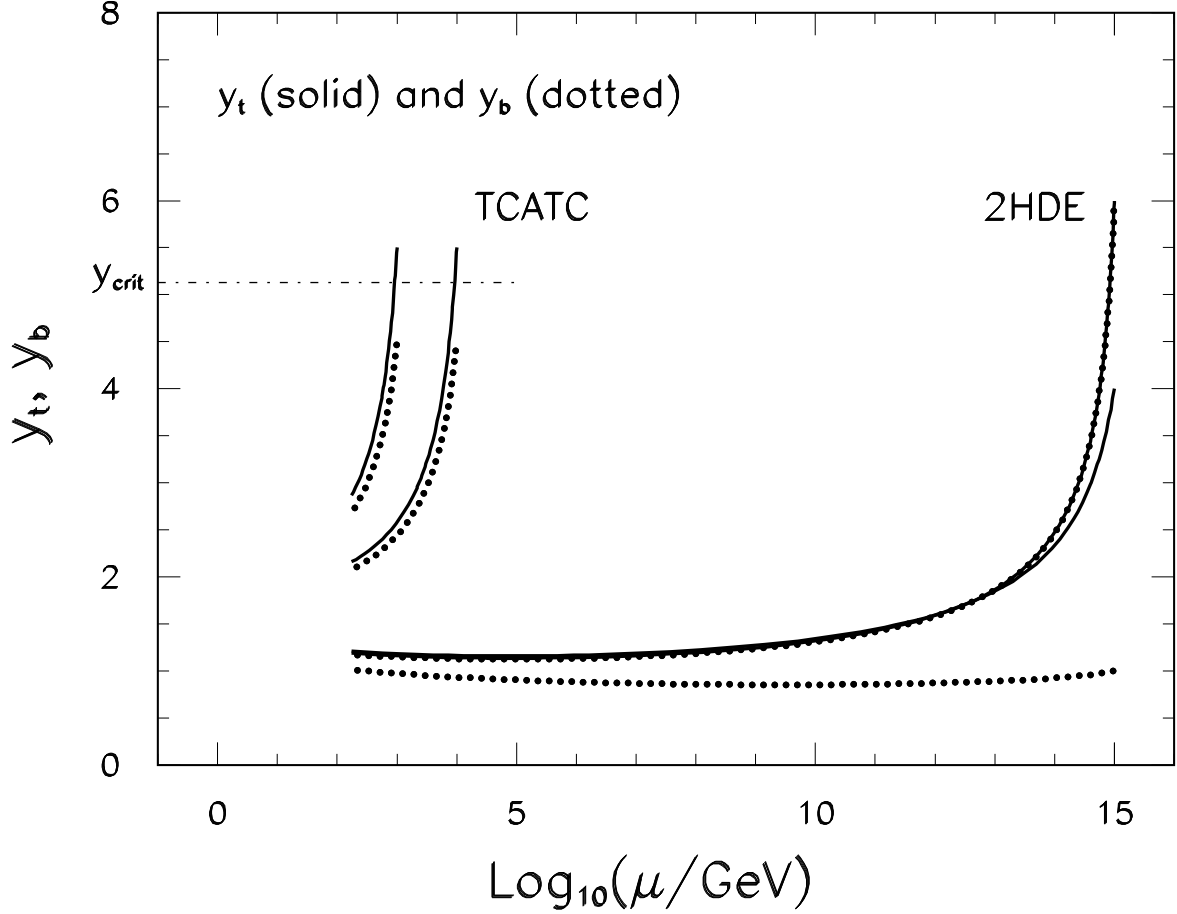}}}
\vskip 0.1 in
\caption{Renormalization group running of $y_t$ and $y_b$ in the
2HDE of top-condensate model and the TCATC model. For the 2HDE of
top-condensate model, two sets of curves are shown: the upper
two curves (solid for $y_t$ and dotted for $y_b$) are for the typical
boundary condition 
$~y_t(10^{15}{\rm GeV})=y_b(10^{15}{\rm GeV})=6 \gg 1~$
and they are too close to be distinguishable; the lower two curves
are for the boundary condition 
$~y_t(10^{15}{\rm GeV})=4, ~y_b(10^{15}{\rm GeV})=1 ~$. In both cases,
$y_t$ and $y_b$ have very similar infrared values, of $O(1)$,
at the weak scale.
}
\label{Fig:TC-RG}
\end{figure}

\newpage
\begin{figure}
\centerline{\hbox{
\psfig{figure=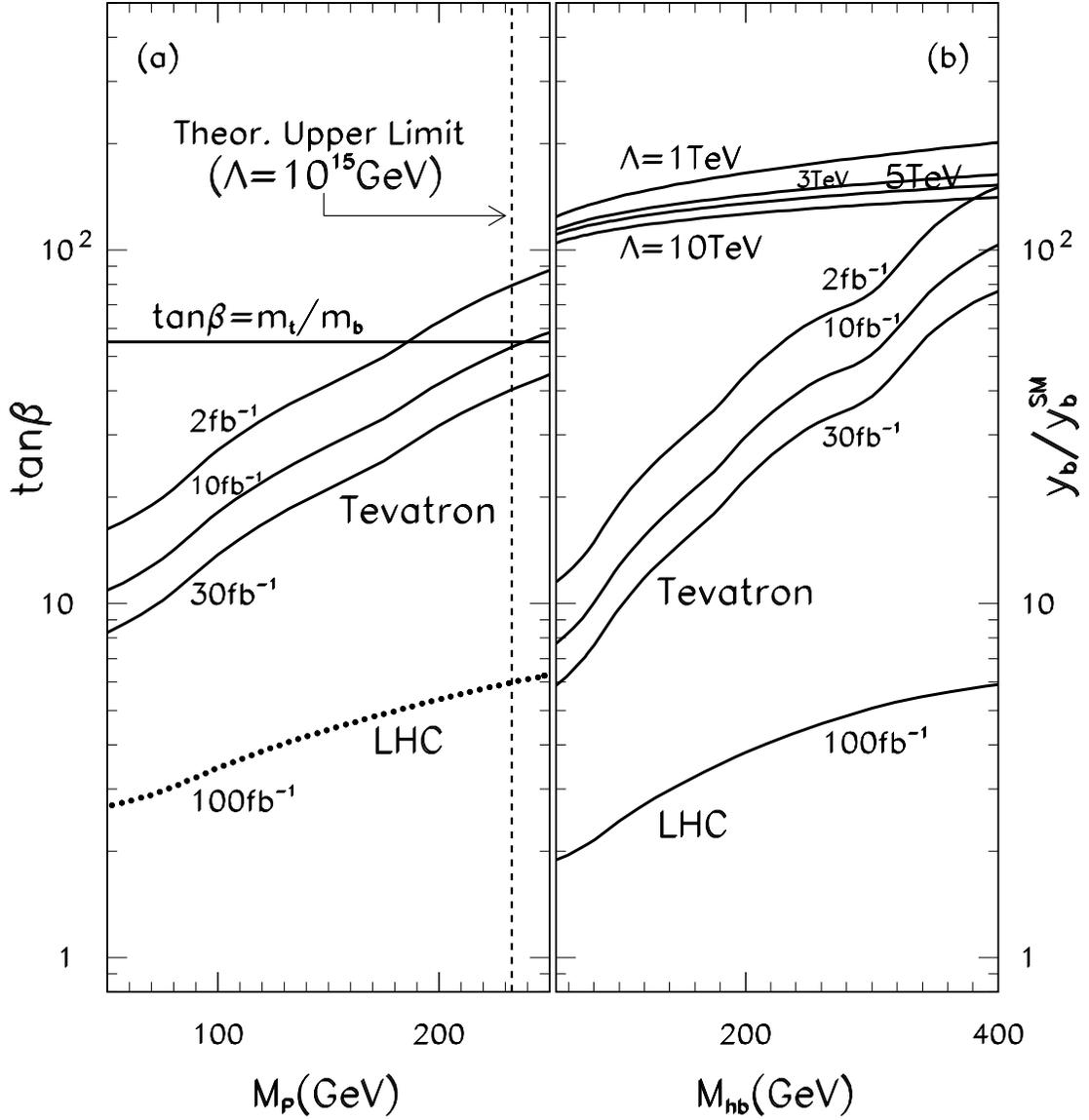,height=18cm}}}
\vskip 0.1 in
\caption{$95\%$~C.L. discovery reach of the Tevatron Run~II 
and the LHC for (a) the 2HDE of top-condensate model and
(b) the TCATC model. Regions above the curves can be discovered.
The top curves in (b) indicate $y_b(\mu =m_t)$ values for
various topcolor breaking scale $\Lambda$, which are based on 
the RG running analysis (cf. Fig.~7).
}
\label{Fig:TC-bound}
\end{figure}

\newpage
\begin{figure}
\centerline{\hbox{
\psfig{figure=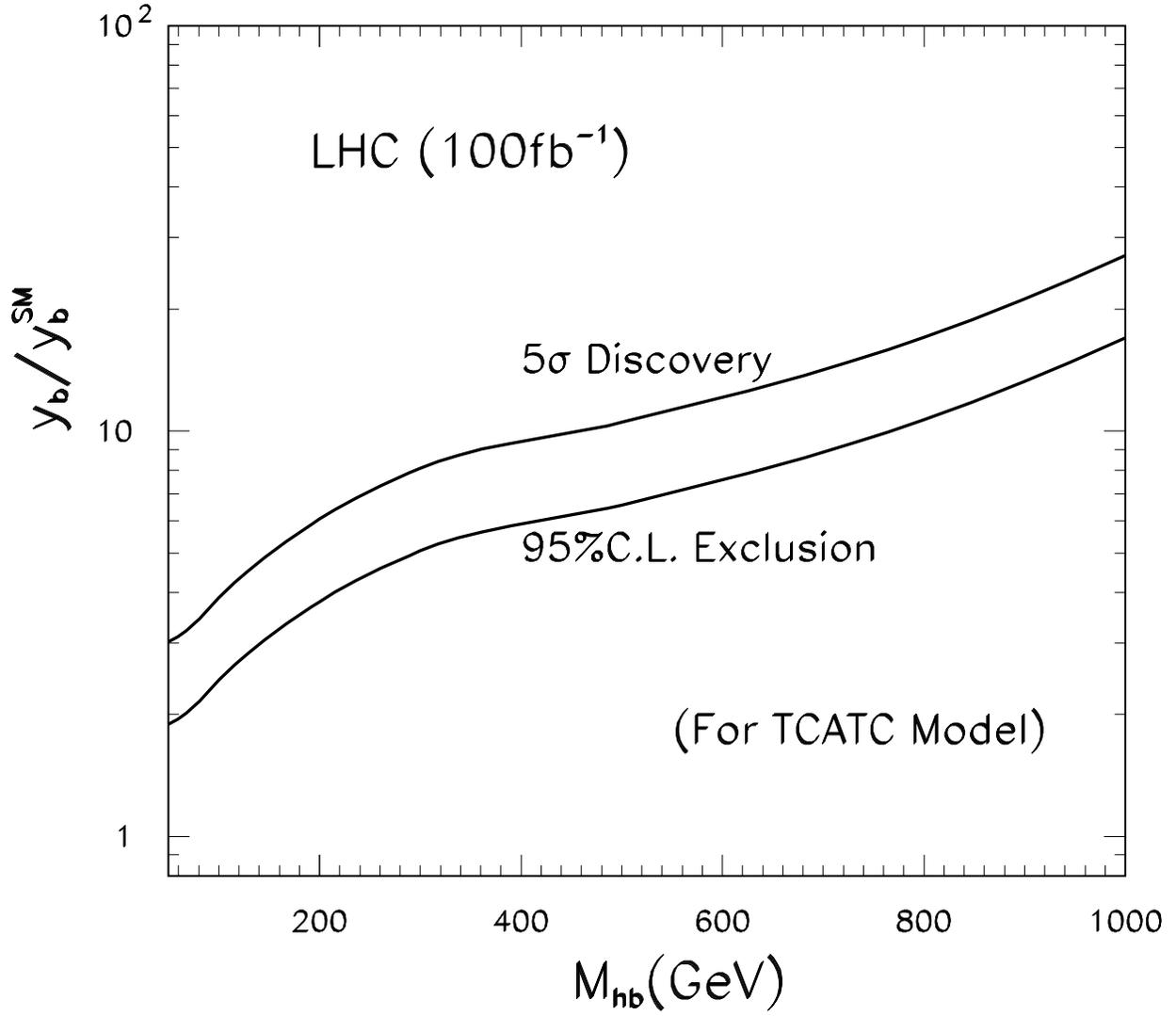}}}
\vskip 0.1 in
\caption{The $5\sigma$ discovery and
$95\%$~C.L. exclusion contours for $y_b(\mu )/y_b^{\rm SM}(\mu )$
as a function of $M_{h_b}$ in the TCATC model,
at the LHC with $100$~fb$^{-1}$ luminosity.
}
\label{Fig:TC-LHCbound}
\end{figure}

\newpage
\begin{figure*}[t]
\vspace*{-1.5cm}
\centerline{\hbox{
\psfig{figure=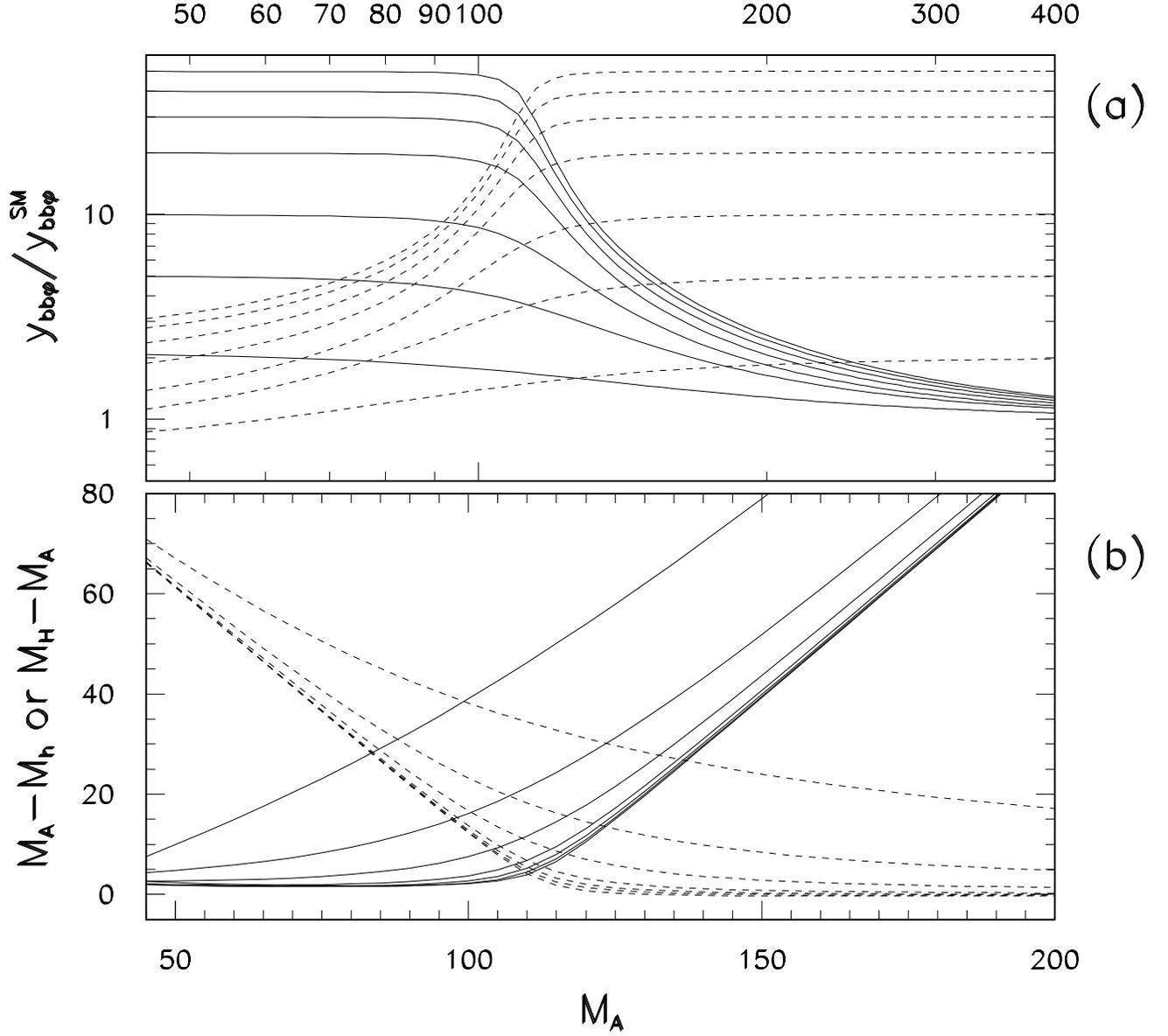,height=7.5in,width=7.5in}}}
\caption{
Bottom Yukawa couplings to the MSSM Higgs bosons and the 
mass differences, $m_A - m_h$ and $m_H - m_A$,
as a function of $m_A$ for $\tan \beta$ values:
2.0, 5.0, 10.0, 20.0, 30.0, 40., 50.0. 
In~(a), $y_{bbh}$ is in solid and $y_{bbH}$ is in dashed,
and $\tan \beta$ decreases from top to bottom curves.
In~(b), $m_A-m_h$ is in solid, $m_H-m_A$ is in dashed, and 
$\tan \beta$ increases from the top to bottom curves. 
Here, all the
SUSY soft-breaking mass parameters are chosen to be 500~GeV.
}
\label{fig:YsMs}
\end{figure*}

\newpage
\begin{figure*}[t]
\centerline{\hbox{
\psfig{figure=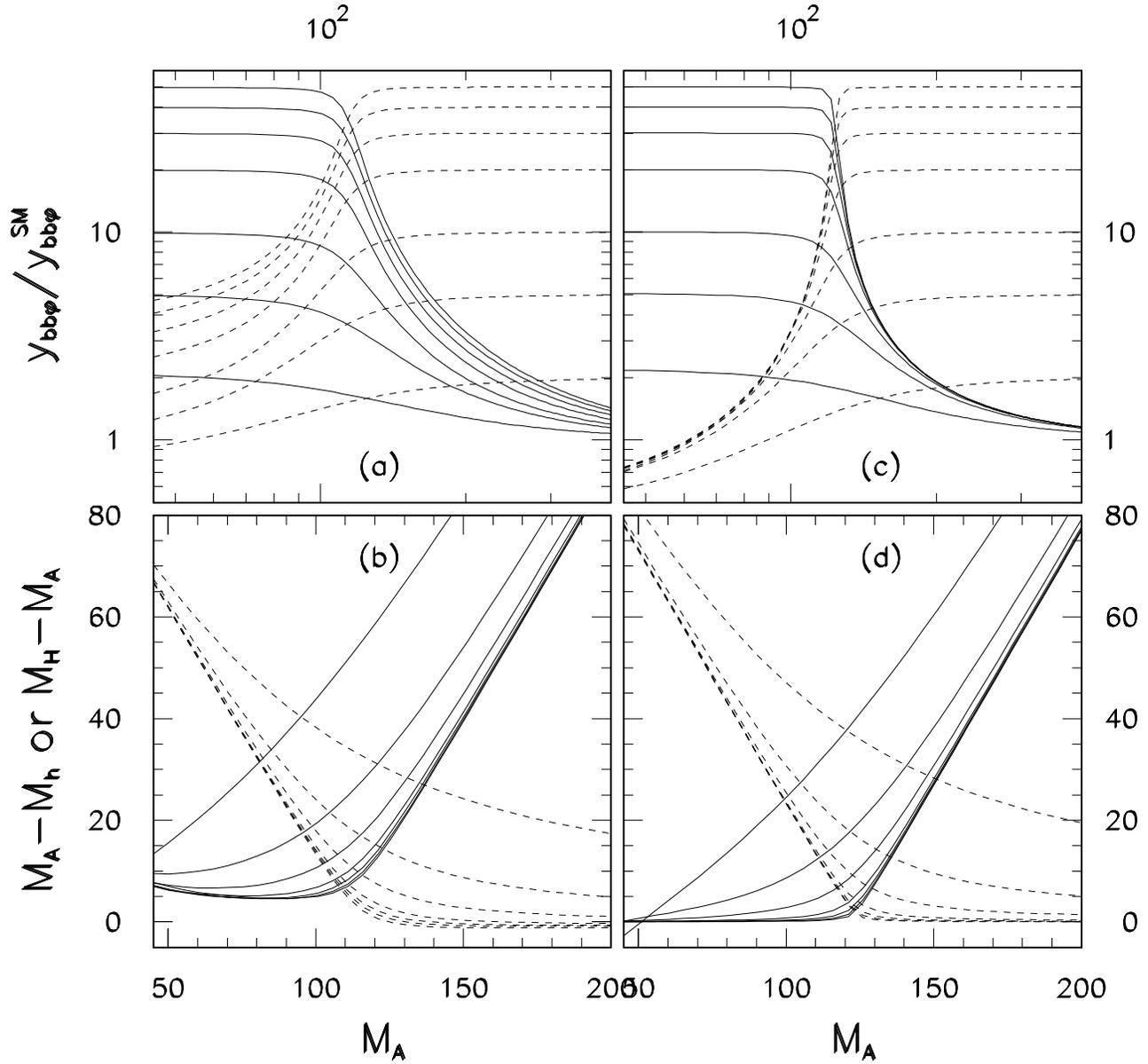,height=7.5in,width=7.5in}}}
\caption{
The same as the previous figure, but in (a)-(b), we change  
the right-handed stop mass to 200~GeV, and in (c)-(d), we use the
``LEP~II Scan A2'' set of SUSY parameters.
}
\label{fig:YsMs2}
\end{figure*}

\newpage
\begin{figure*}[t]
\begin{center}
\begin{tabular}{cc} 
\psfig{figure=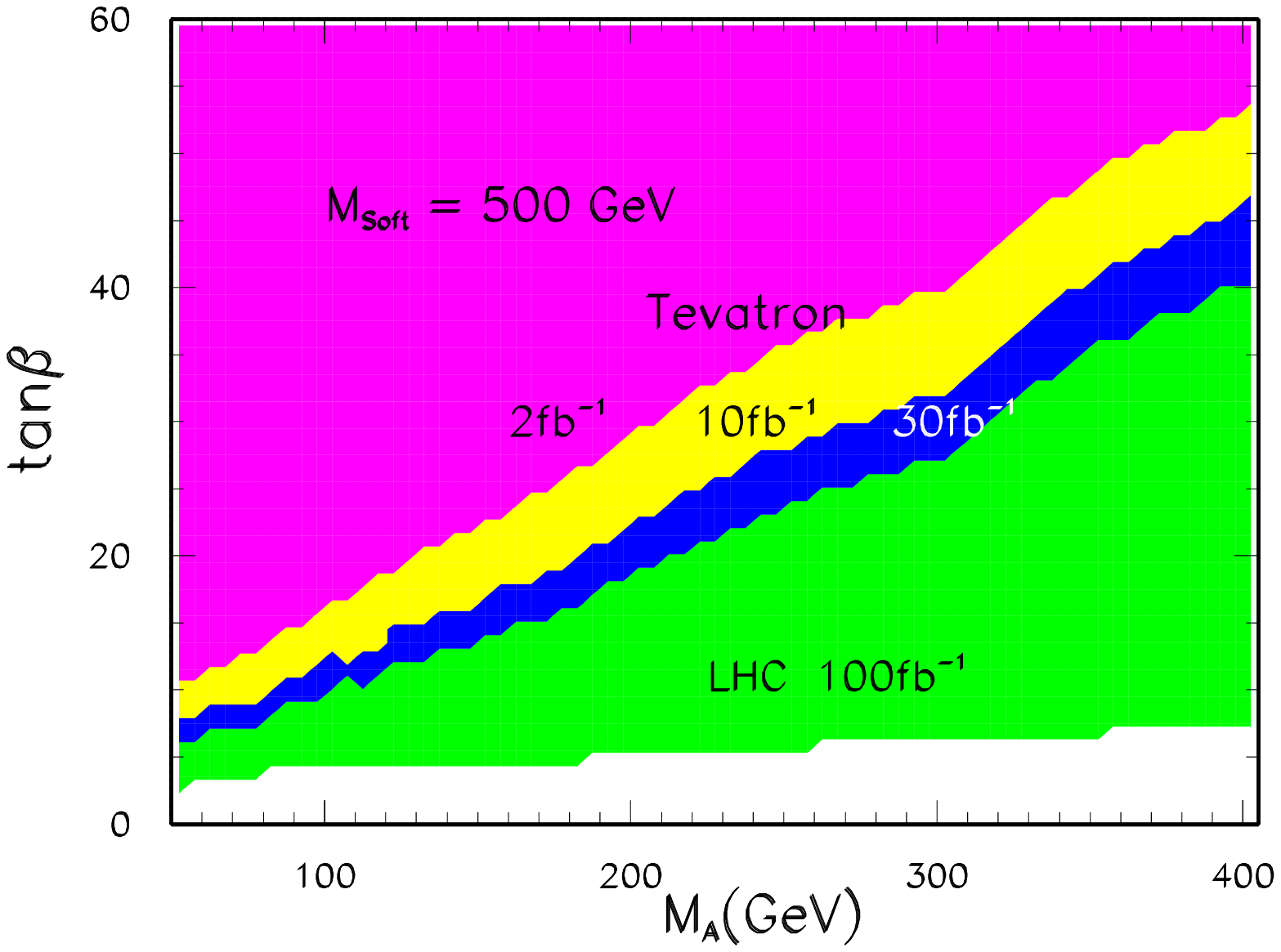,height=3.8in} 
\\[-3.4in]\hskip 4.7in (a)\\[3.1in] \\[-2.0cm] 
\psfig{figure=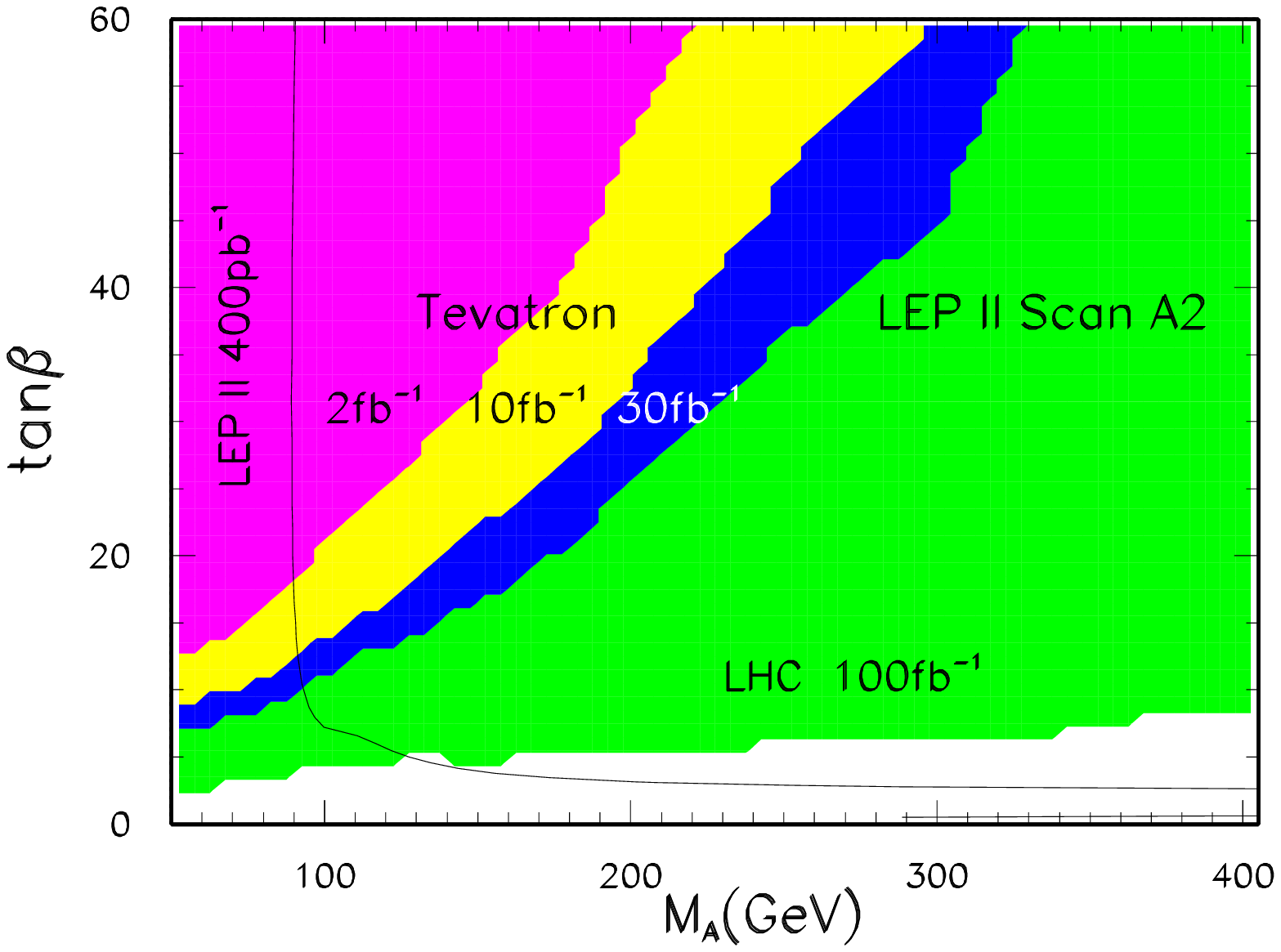,height=3.8in} 
\\[-3.4in]\hskip 4.7in (b)\\[3.1in]
\end{tabular}
\end{center}
\vspace*{-0.5cm}
\caption{
$95\%$~C.L. exclusion contours in the 
$m_A$-$\tan\beta$ plane of the MSSM.
The areas above the four boundaries are excluded for the Tevatron Run~II
with the indicated luminosities,
and for the LHC with an integrated luminosity of 100 fb$^{-1}$.
The soft SUSY breaking parameters were chosen uniformly to be 500 GeV
in Fig.~(a), while the inputs of the ``LEP~II Scan~A2'' are used for
the Fig.~(b) in which LEP~II excludes the left area of the solid curve. 
}
\label{fig:Exclusion}
\end{figure*}

\newpage
\begin{figure*}[t]
\centerline{\hbox{
\psfig{figure=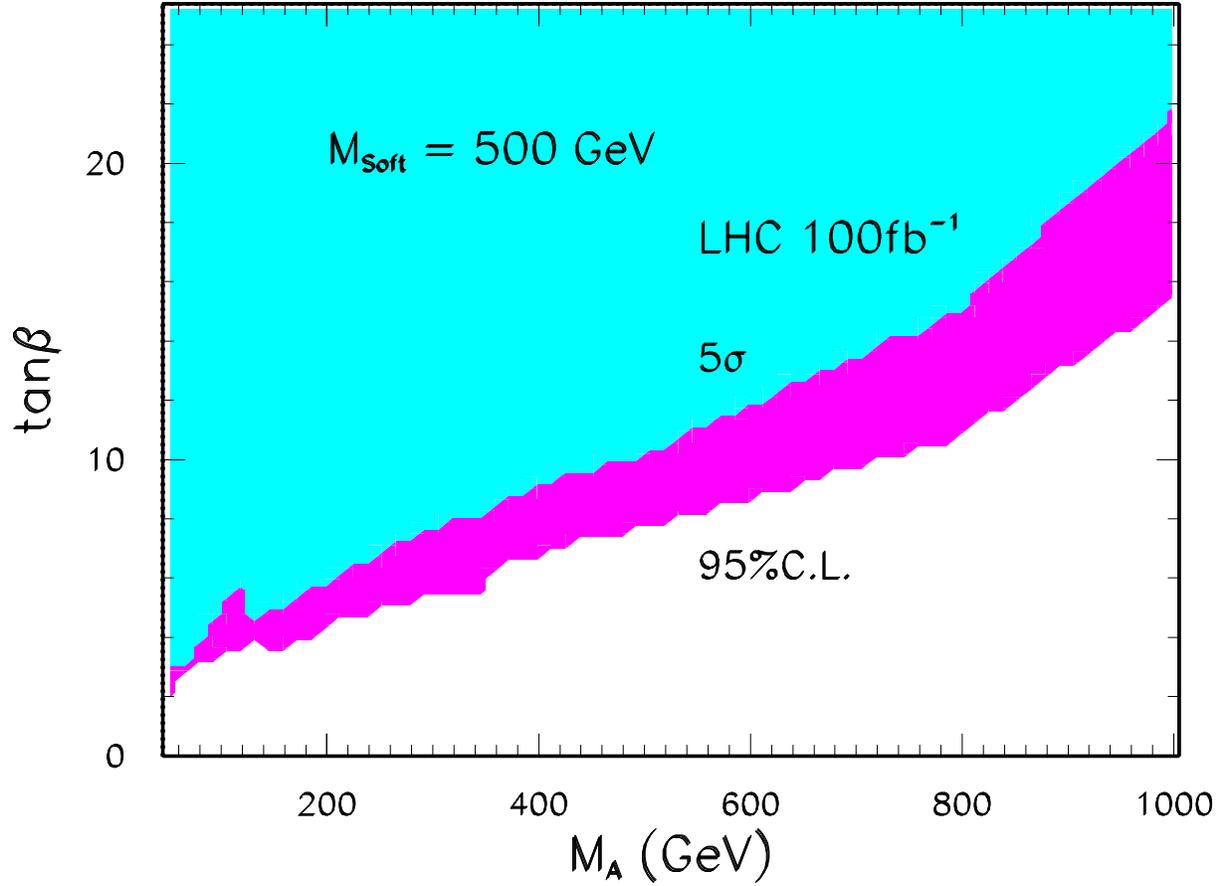}}}
\vskip 0.1 in
\caption{
Discovery and exclusion contours in the 
$m_A$-$\tan\beta$ plane of the MSSM 
for the LHC with an integrated luminosity of 100 fb$^{-1}$.
The area above the lower boundary is excluded at $95\%$~C.L., while
the upper boundary is the $5 \sigma$ discovery contour.
The soft SUSY breaking parameters were chosen uniformly to be 500 GeV.
}
\label{fig:ExclusionLHC}
\end{figure*}

\newpage
\begin{figure*}[t]
\begin{center}
\begin{tabular}{cc} 
\psfig{figure=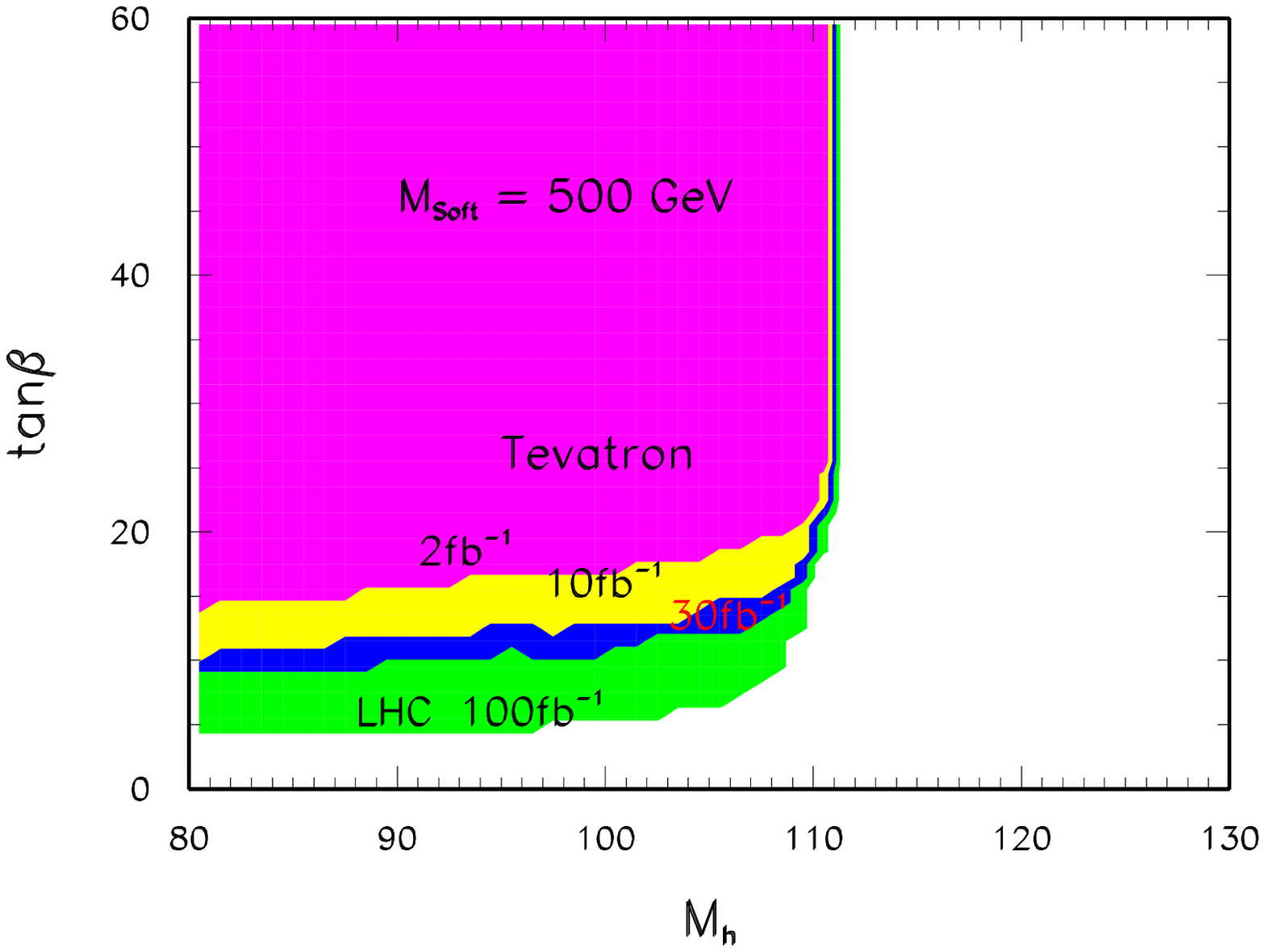,height=3.8in} 
\\[-3.4in]\hskip 4.7in (a)\\[3.1in] \\[-1.6cm] 
\psfig{figure=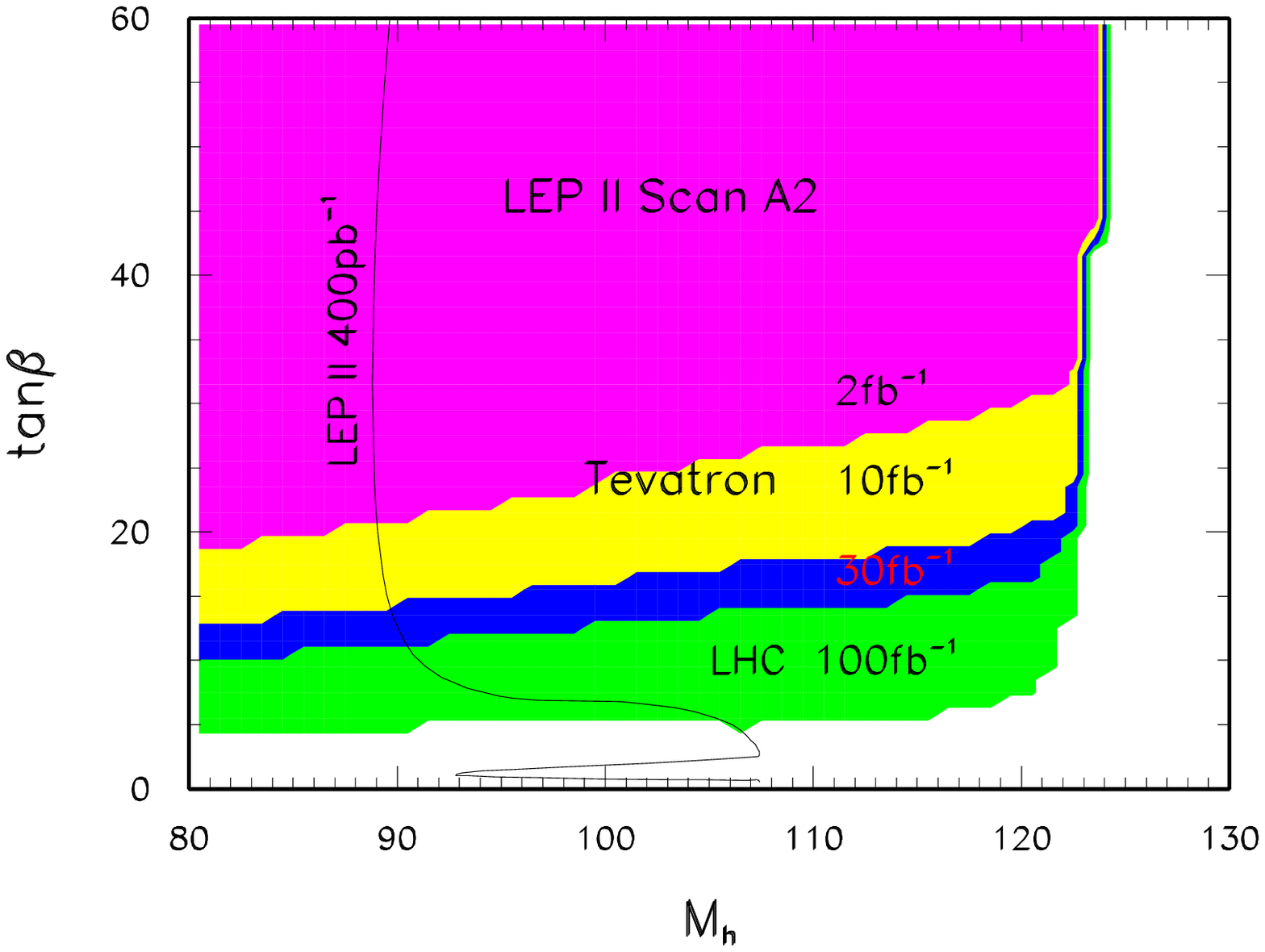,height=3.8in} 
\\[-3.4in]\hskip 4.7in (b)\\[3.1in]
\end{tabular}
\end{center}
\vspace*{-0.5cm}
\caption{
$95\%$~C.L. exclusion contours in the $m_h$-$\tan\beta$ plane of the MSSM.
The areas above the four boundaries are excluded for the Tevatron Run~II
with the indicated luminosities,
and for the LHC with an integrated luminosity of 100 fb$^{-1}$.
LEP~II can exclude the area on the left-hand side of the solid 
curve in the lower plot. 
}
\label{fig:Exclusion2}
\end{figure*}

\newpage
\begin{figure*}[t]
\begin{center}
\begin{tabular}{cc} 
\psfig{figure=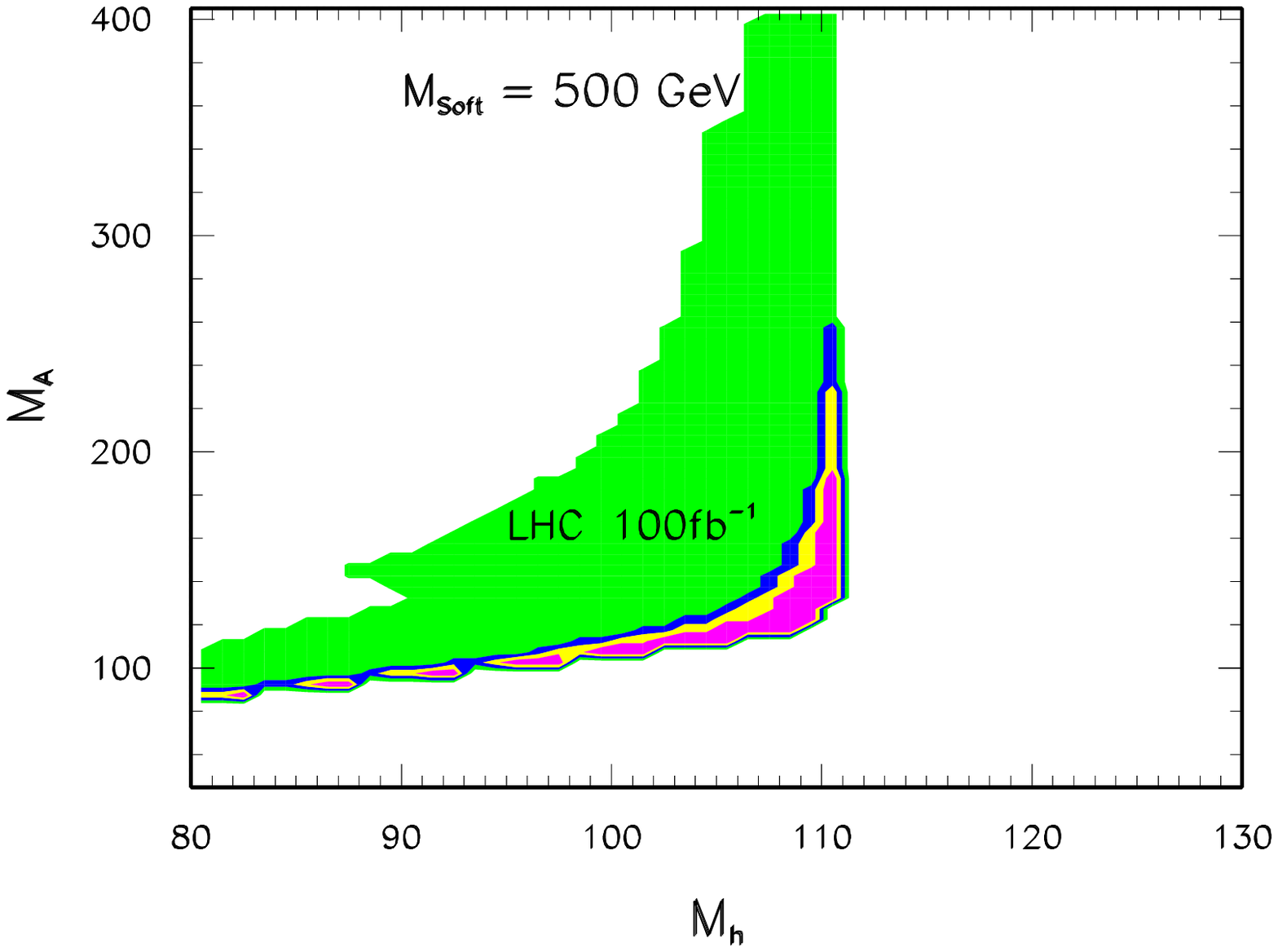,height=3.8in} 
\\[-3.4in]\hskip 4.7in (a)\\[3.1in] \\[-1.6cm] 
\psfig{figure=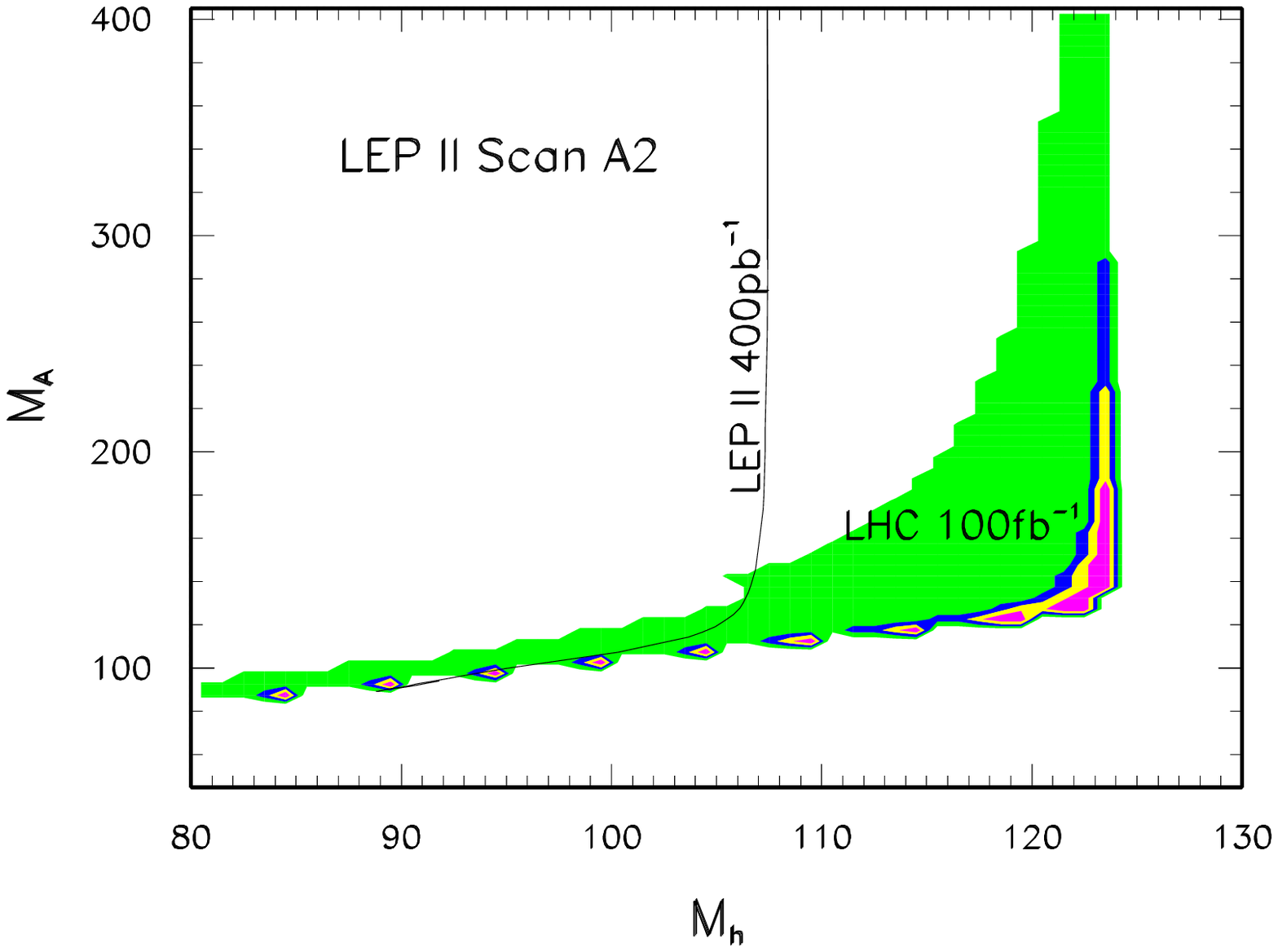,height=3.8in} 
\\[-3.4in]\hskip 4.7in (b)\\[3.1in]
\end{tabular}
\end{center}
\vspace*{-0.5cm}
\caption{
$95\%$ ~C.L. exclusion contours in the $m_A$-$m_h$ plane of the MSSM.
The shaded areas indicate the excluded regions for the Tevatron Run~II
with integrated luminosities, 2, 10, 30 fb$^{-1}$,
and for the LHC with an integrated luminosity of 100 fb$^{-1}$,
as those in the previous figures.
LEP~II can exclude the area on the left-hand side of the
solid curve in the lower plot. 
}
\label{fig:Exclusion3}
\end{figure*}

\newpage
\begin{figure*}[t]
\centerline{\hbox{
\psfig{figure=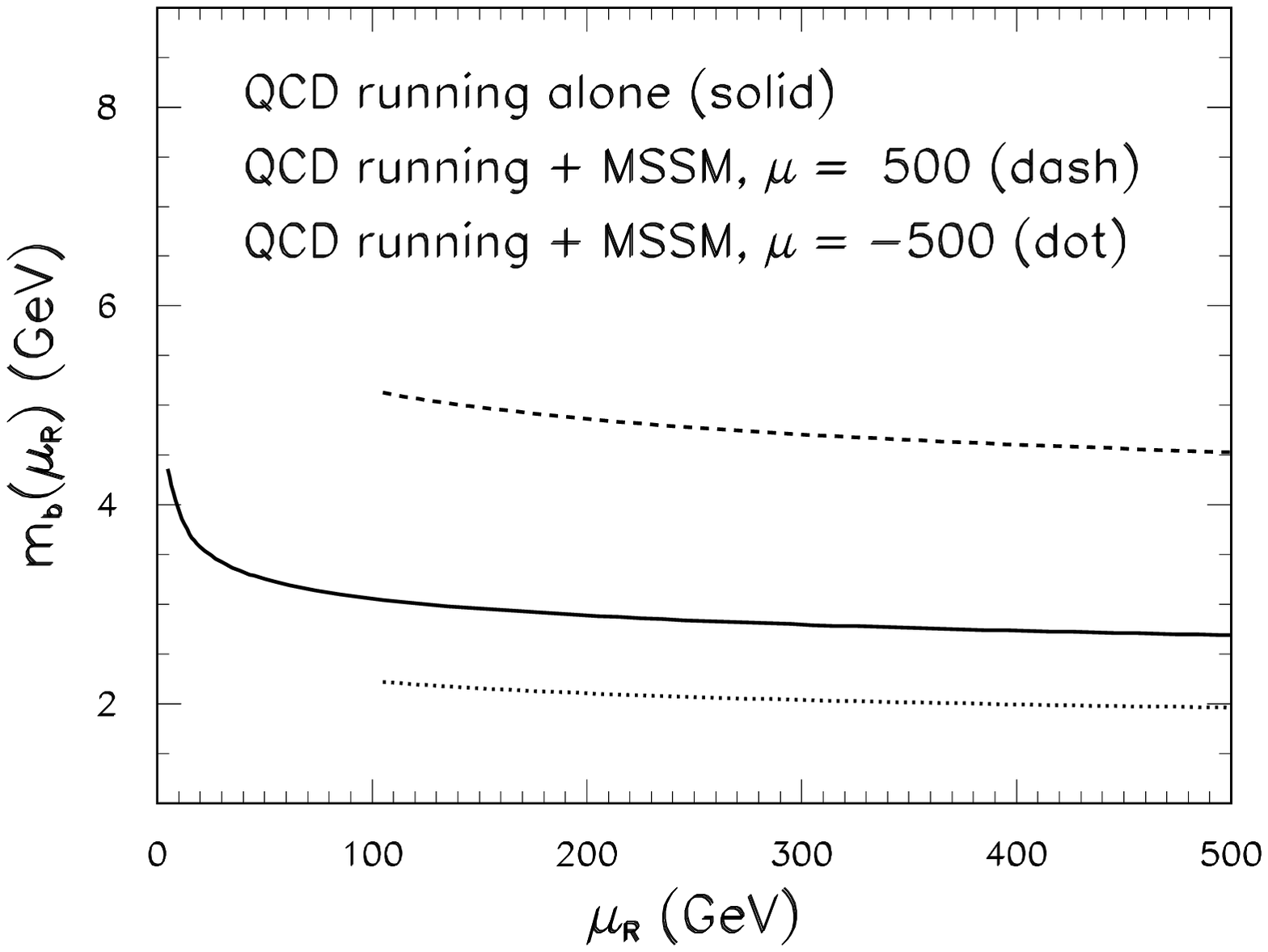}}}
\vskip 0.1 in
\caption{
The running of the bottom quark mass $m_b(\mu_R)$ 
as a function of the renormalization scale $\mu_R$.  
The solid curve shows the QCD evolution alone. 
The dashed curve further includes the supersymmetric corrections to the 
``effective'' running mass, for $\tan \beta = 30$. 
All soft SUSY breaking parameters have been fixed as $500$~GeV.
The dotted curve includes the SUSY corrections but with the sign
of the Higgs-mixing parameter $\mu$ flipped.
}
\label{fig:mbRun2L}
\end{figure*}


\begin{thebibliography}{99}

\bibitem{hhunt} 
For reviews, see {\it e.g.}, 
G. Kane, {\it Perspective on Higgs Physics,} 
2nd edition, World Scientific, 1998; 
T.L. Barklow, {\it et al,} 
{\it Strong Electroweak Symmetry Breaking,}
hep-ph/9704217, in ``New Directions for High Energy Physics'',
Snowmass, CO, June, 1996; 
H. Haber, {\it Future Directions in Higgs Phenomenology,}
hep-ph/9703381, in ``The Higgs Puzzle: 
What can we learn from LEP2, LHC, 
NLC and FMC'', ed. B.A. Kniehl, (World Scientific, 1997); 
J.F. Gunion, {\it Detecting and Studying
Higgs Bosons,} hep-ph/9705282;
J.F. Gunion, H.E. Haber, G. Kane and S. Dawson, 
{\it The Higgs Hunter's Guide,} Addison-Wesley Pub., 
1996 (2nd edition). 

\bibitem{topmass}
F.~Abe, {\it et al.}, 
Phys. Rev. Lett. {\bf 74}, 2626 (1995); {\bf 77}, 438 (1996); 
S.~Adachi, {\it et al.}, Phys. Rev. Lett. {\bf 74}, 2632 (1995); 
P. Giromini, Proceedings of the Lepton-Photon Symposium, 
Hamburg, August 1997. 

\bibitem{topCrev}
For a review, G.~Cvetic,  Rev. Mod. Phys. (to appear), 
hep-ph/9702381, and references therein.

\bibitem{SUSY-rad}
L. Ibanez, Nucl. Phys. B{\bf 218}, 514 (1983); 
J. Ellis, D. Nanopolous and K. Tamvakis, Phys. Lett. 
B{\bf 121}, 123 (1983); 
L. Alvare-Gaume, J. Polchinski and M. Wise, Nucl. Phys. 
B{\bf 221}, 495 (1983).

\bibitem{IRQFP}
C.T.~Hill,  Phys. Rev. D{\bf 24}, 691 (1981).

\bibitem{mssm} 
For examples,
J. Gunion and H. Haber, Nucl. Phys. B{\bf 272}, 1 (1986);
H. Haber, hep-ph/9306207, hep-ph/9709450 and hep-ph/9703391; 
S.Dawson, hep-ph/9612229; S.P. Martin, hep-ph/9709356. 

\bibitem{bbbb_prl} 
J.L. Diaz-Cruz, H.-J. He, T. Tait, and C.-P. Yuan,
Phys. Rev. Lett. {\bf 80}, 4641 (1998), hep-ph/9802294.

\bibitem{dai}
J. Dai, J. Gunion, and R. Vega, Phys. Lett. B{\bf 345}, 29 (1995); 
 B{\bf 387}, 801 (1996).

\bibitem{tt-2HDM}
M.A.~Luty, Phys. Rev. D{\bf 41}, 2893 (1990); 
M.~Suzuki, Phys. Rev. D{\bf 41}, 3457 (1990).

\bibitem{topcolor}
C.T.~Hill, Phys. Lett. B{\bf 345}, 483 (1995);
C.T.~Hill, hep-ph/9702320, in proceedings
of {\it Strongly Coupled Gauge Theories,} November, 1996, Nagoya, Japan.

\bibitem{tt-lindner}
M.~Lindner, hep-ph/9704362; 
E.~Akhmedov, M.~Lindner, E.~Schnapka, and J.W.F.~Valle, 
Phys. Rev. D{\bf 53}, 2752 (1996); Phys. Lett. B{\bf 368}, 270 (1996); 
M.~Lindner and E.~Schnapka, hep-ph/9712489, TUM-HEP-293/97, LBNL-41081.

\bibitem{BHL}
W.A.~Bardeen, C.T.~Hill, and M.~Lindner, Phys. 
Rev. D{\bf 41}, 1647 (1990).

\bibitem{susyrev} 
H.P. Nilles, Phys. Rep. {\bf 110}, 1 (1984); 
H. Haber and G.L. Kane, Phys. Rep. {\bf 117}, 75 (1985).

\bibitem{GMSB}
For a recent review,
G.F. Giudice and R. Rattazzi, CERN-TH-97 and hep-ph/9801271.

\bibitem{dicus}
J.C. Collins and W.-K. Tung, Nucl. Phys. B{\bf 278}, 934 (1986);
D. A. Dicus and S. Willenbrock, Phys. Rev. D{\bf 39}, 751 (1989).

\bibitem{bbbbbackground}
V. Barger, A. L. Stange, and R. J. N. Phillips, Phys. Rev. D{\bf 44},
1987 (1991);
J. F. Gunion and Z. Kunszt, Phys. Lett. B{\bf 159}, 167 (1985);
{\bf 176}, 477 (1986).

\bibitem{madgraph}
W.F. Long and T. Stelzer, Comput. Phys. Commun. {\bf 81}, 357 (1994).

\bibitem{kfac}
M. Mangano, P. Nason, and G. Ridolfi, Nucl. Phys.
B{\bf 373}, 295 (1992).

\bibitem{dawsonreina}
S. Dawson and L. Reina, Phys. Rev. D{\bf 57}, 5851 (1998).

\bibitem{cteq}
CTEQ Collaboration: H. Lai, J. Huston, S. Kuhlmann,
F. Olness, J. Owens, D. Soper, W.-K. Tung, and H. Weerts,
Phys. Rev. D{\bf 55}, 1280 (1997).

\bibitem{tev2000}
D. Amidei and R. Brock, ``Report of the $TeV 2000$ Study
Group on Future Electroweak Physics at the Tevatron'', 1995.

\bibitem{froidevaux}
E. Richter-Was and D. Froidevaux, Z. Phys. C{\bf 76}, 665 (1997).

\bibitem{pythia}
T. Sjostrand, Comput. Phys. Commun. {\bf 82}, 74 (1994);
S. Mrenna, Comput. Phys. Commun. {\bf 101}, 232 (1997).

\bibitem{mrrs}
A. Martin, R. Roberts, M. G. Ryskin,
and W. J. Stirling, ``Consistent Treatment of Charm Evolution
in Deep Inelastic Scattering'', hep-ph/9612449.

\bibitem{topcolor2}
K.~Lane and E.~Eichten, Phys. Lett.  B{\bf 352}, 382 (1995).

\bibitem{topcolor3}
R.S.~Chivukula and H.~Georgi, hep-ph/9805478 and hep-ph/9806289.

\bibitem{Nambu}
Y.~Nambu, in {\it New Theories in in Physics,} Proceedings of the XI
International Symposium on Elementary Particle Physics, pp.1-10,
1988, eds. Z.~Ajduk et al, (World Scientific).
 
\bibitem{top-seesaw}
B.A.~Dobrescu and C.T.~Hill, 
Phys. Rev. Lett. {\bf 81}, 2634 (1998).
R.S.~Chivukula, B.A.~Dobrescu, H.~Georgi and C.T.~Hill,
hep-ph/9809470.


\bibitem{susy-topc}
C.D.~Froggatt, I.G.~Knowles and R.G.~Moorhouse, 
Phys. Lett. B{\bf 298}, 356 (1993).

\bibitem{NJL}
Y.~Nambu and G.~Jona-Lasinio, Phys. Rev. {\bf 122}, 345 (1961).

\bibitem{auxiliary}
D. Bjorken, Ann. Phys. {\bf 24}, 174 (1963);
T. Equchi, Phys. Rev. D{\bf 14}, 2755 (1976).

\bibitem{RGE}
C.T.~Hill, C.~Leung, and S.~Rao, Nucl. Phys. B{\bf 262}, 517 (1985).

\bibitem{Barger}
For example, V.~Barger, M.S.~Berger, and P.~Ohmann,
Phys. Rev. D{\bf 47}, 1093 (1993), and references therein.

\bibitem{PDG}
R.M.~Barnett {\it et al.}, Phys. Rev. D{\bf 54}, 1 (1996).

\bibitem{hdecay} 
 A. Djouadi, J. Kalinowski and M. Spira, 
Comput. Phys. Commun. {\bf 108}, 56 (1998).

\bibitem{PS}
H.~Pagels and S.~Stokar, Phys. Rev. D{\bf 20}, 2947 (1979).

\bibitem{ETC0}
S.~Weinberg, Phys. Rev. D{\bf 19}, 1277 (1979);
L.~Susskind, Phys. Rev. D{\bf 20}, 2619 (1979);
S.~Dimopoulos and L.~Susskind, Nucl. Phys. B{\bf 155}, 237 (1979);
E.~Eichten and K.~Lane, Phys. Lett. B{\bf 90}, 125 (1980).

\bibitem{ETC}
B.~Holdom, 
Phys. Rev. D{\bf 24}, 1441 (1981); Phys. Lett. B{\bf 150}, 301 (1985);
T.~Appelquist, D.~Karabali, and L.C.R.~Wijewardhana,
Phys. Rev. Lett. {\bf 57}, 957 (1986);
K.~Yamawaki, M.~Bando, and K.~Matumoto,
Phys. Rev. Lett. {\bf 56}, 1335 (1986).

\bibitem{topcolor-more}
D.~Kominis, Phys. Lett. B{\bf 345}, 483 (1995);
G.~Buchalla, G.~Burdman, C.T.~Hill, and D.~Kominis,
Phys. Rev. D{\bf 53}, 5185 (1996);
J.D.~Wells, Phys. Rev. D{\bf 56}, 1504 (1997).

\bibitem{lindner}
M.~Lindner, private communications.

\bibitem{himrc}
Y. Okada {\it et al.,} Prog. Theor. Phys. Lett. {\bf 85}, 1 (1991);
H. Haber and R. Hempfling, Phys. Rev. Lett. {\bf 66}, 1815 (1991);
J. Ellis {\it et al.,} Phys. Lett. B{\bf 257}, 83 (1991).

\bibitem{madison-LEP2}
C.~Rembser, {\it Recent Results on LEP~II,} plenary talk 
at International Symposium on `` Frontiers of Phenomenology 
from Non-perturbative QCD to New Physics '', March~23-26, 1998,
Madison, Wisconsin.

\bibitem{FNAL-LEP2} 
N.~Konstantinidis, {\it Higgs searches at LEP}, plenary talk 
at the Workshop on `` Physics at Run II -- Supersymmetry/Higgs '', 
First General Meeting: 14-16 May, 1998, 
Fermi National Accelerator Laboratory, Batavia, Illinois.

\bibitem{dressetal} 
M. Drees, M.~Guchait, and P.~Roy, 
Phys. Rev. Lett. {\bf 80}, 2047 (1998), 
                 {\bf 81}, 2394(E) (1998).

\bibitem{stev} 
S.~Mrenna, talk given at the Higgs Working Group
meeting of the Run~II Workshop,
Fermilab, May~13-16, 1998;
M.~Carena, S.~Mrenna and C.E.W. Wagner, to appear.

\bibitem{spira}  
M.~Spira, private communication.

\bibitem{mbrun2}
R.~Hempfling, Phys. Rev.  D{\bf 49}, 6168 (1994);
L.J.~Hall, R.~Rattazzi, and U.~Sarid,  
Phys. Rev.  D{\bf 50}, 7048 (1994);
M.~Carena, M.~Olechowski, S.~Pokorski, and C.E.M. Wagner,  
Nucl. Phys. B{\bf 426}, 269 (1994).

\bibitem{mbrun}
J. Bagger, K. Matchev, and D. Pierce, hep-ph/9503422,
in Proceedings of {\it Beyond the Standard Model IV,} p.363,
Lake Tahoe, CA, Dec~13-18, 1994;
D.M. Pierce, J.A. Bagger, K. Matchev, and R.J. Zhang,  
Nucl. Phys. B{\bf 491}, 3 (1997), hep-ph/9606211,  
and references therein.

\bibitem{mbrun3}
J.A. Coarasa, R.A. Jimenez and J. Sola, 
Phys. Lett. B{\bf 389}, 312 (1996); 
R.A. Jimenez and J. Sola, {\it ibid.} B{\bf 389}, 52 (1996).

\bibitem{CWZ}
J.~Collins, F.~Wilczek, and A.~Zee, Phys. Rev. D{\bf 18}, 242 (1978).

\bibitem{kanet} 
G.L. Kane, hep-ph/9705382 and hep-ph/9709318;
J.L. Feng {\it et al.,}  Phys. Rev. D{\bf 52}, 1419 (1995).

\bibitem{sotengut} 
See, for instance: 
L.J.~Hall, R.~Rattazzi, and U.~Sarid, Phys. Rev. D{\bf 50}, 7048 (1994);
S.~Dimopoulos, L.J.~Hall, and S.~Raby, 
Phys. Rev. Lett. {\bf 68}, 1984 (1992);
G. Lazarides and C. Panagiotakopoulos, hep-ph/9407285.

\bibitem{irfixedp} 
See, for instance: 
 C.D. Froggatt, R.G. Moorhouse, I.G. Knowles,  
Phys. Lett. B{\bf 298}, 356 (1993);
B.~Schrempp and M.~Wimmer, 
Prog. Part. $\&$ Nucl. Phys. {\bf 37}, 1 (1996);
and references therein. 

\bibitem{moremssm} A. Cohen, D. Kaplan and A. Nelson,
Phys. Lett. B{\bf 388}, 588 (1996).

\bibitem{sugrarev} For a review, 
R. Arnowitt and P. Nath, hep-ph/9708254.

\bibitem{dressnoji}
M. Drees and M. Nojiri, Nucl. Phys B{\bf 369}, 54 (1992).

\bibitem{sugrtbg}
H. Baer {\it et al.,} Phys. Rev. Lett. {\bf 79}, 986 (1997).

\bibitem{dinetal} 
M. Dine, A. Nelson, Y.~Nir, and Y.~Shirman,
Phys. Rev. D{\bf 53}, 2658 (1996). 

\bibitem{gmmrev} 
For a review, C. Kolda, hep-ph/9707450.

\bibitem{babuetal} 
K. Babu, C. Kolda,  and F. Wilczek, 
Phys. Rev. Lett. {\bf 77}, 3070 (1996);
S.~Dimopoulos, M.~Dine, S.~Raby, and  S.~Thomas, 
Phys. Rev. Lett. {\bf 76}, 3494 (1996).

\bibitem{baggetal} 
J.A. Bagger, K. Matchev, D.M. Pierce, and R.J. Zhang, 
Phys. Rev. D{\bf 55}, 3188 (1997).

\bibitem{baertal} 
H.~Baer, M.~Brhlik, C.~Chen, and X.~Tata,
Phys. Rev. D{\bf 55}, 4463 (1997).

\bibitem{ratsarid} 
R. Rattazzi and U. Sarid, Nucl. Phys. B{\bf 501} 297 (1997);
E. Gabrielli and U. Sarid, Phys. Rev. Lett. {\bf 79}, 4752 (1997).

\bibitem{borzumati} 
F. Borzumatti, hep-ph/9702307

\bibitem{himlim} 
ALEPH Collaboration, R. Barate {\it et al.,} Phys. Lett. 
B{\bf 412}, 173 (1997); 
DELPHI Collaboration, P. Abreu {\it et al.,} 
Eur. Phys. J. C{\bf 2}, 1 (1998);
L3 Collaboration, M. Acciarri {\it et al.,} CERN-EP-98-072 
(submitted to Phys. Lett. B); 
OPAL Collaboration, K. Ackerstaff {\it et al.,} 
e-Print: hep-ex/9803019 (submitted to Eur. Phys. J. C); 
J.F. Gunion {\it et al.,} 1996 DPF/DPS summer study, hep-ph/9703330.

\bibitem{kunstz}
Z. Kunszt and F. Zwirner, Nucl. Phys. B{\bf 385}, 3 (1992).

\bibitem{gunorr}
J.L.Diaz-Cruz and O. Sampayo, J. Mod. Phys. A{\bf 8}, 4339 (1993);
J. Gunion and L. Orr, Phys. Rev D{\bf 46}, 2052 (1994);
H. Baer, B.W. Harris, X. Tata , hep-ph/9807262.

\bibitem{eehiggs} 
See, for instance: J.F. Gunion {\it et al.}, 1996 DPF/DPS 
summer study, hep-ph/9703330; J.F~Gunion, hep-ph/9801417.

\bibitem{eehiggs2} 
H.~Murayama and M.E.Peskin,  Ann. Rev. Nucl. Part. Sci.
{\bf 46}, 533 (1997) and hep-ex/9606003;
E. Accomando {\it et al.,} Phys. Rep. {\bf 299}, 1 (1998)
and hep-ph/9705442.

\bibitem{cdfwh}
F. Abe {\it et al.}, The CDF Collaboration,
Fermilab-Pub-98/252-E.

\bibitem{run2btag}
The CDF II Collaboration, 
{\it The CDF II Detector Technical Design Reprot}, Fermilab-Pub-96/390-E.

\end{thebibliography}
\end{document}

\\
Title:     Probing Higgs Bosons with Large Bottom Yukawa Coupling 
                          at Hadron Colliders 
Authors:   C. Balazs, J.L. Diaz-Cruz, H.-J. He, T. Tait, C.-P. Yuan 
Comments:  Version to be published in Phys. Rev. D, RevTex, 52 pages,   
           including 16 EPS figures, typos corrected plus minor refinements
           for discussions
Report-no: MSUHEP-80515, FermiLab-Pub-98/182-T, ANL-HEP-PR-98-55
\\
The small mass of the bottom quark, relative to its weak isospin partner, the 
top quark, makes the bottom an effective probe of new physics in Higgs and top 
sectors. We study the Higgs boson production associated with bottom quarks,
ppbar/pp to phi+b+bbar to 4b, at the Fermilab Tevatron and the CERN LHC. We 
find that strong and model-independent constraints on the size of the 
phi-b-bbar coupling can be obtained for a wide range of Higgs boson masses. 
Their implications for the composite Higgs models with strong dynamics 
associated with the third family quarks (such as the top-condensate/topcolor 
models with naturally large bottom Yukawa couplings), and for the 
supersymmetric models with large tan(beta), are analyzed.  We conclude that 
the Tevatron and the LHC can put stringent bounds on these models, if the 
phi+b+bbar signal is not found. 
\\